\documentclass{article}

\usepackage{arxiv}

\usepackage[utf8]{inputenc} % allow utf-8 input
\usepackage[T1]{fontenc}    % use 8-bit T1 fonts
\usepackage{hyperref}       % hyperlinks
\usepackage{url}            % simple URL typesetting
\usepackage{booktabs}       % professional-quality tables
\usepackage{amsfonts}       % blackboard math symbols
\usepackage{nicefrac}       % compact symbols for 1/2, etc.
\usepackage{microtype}      % microtypography
\usepackage{lipsum}		% Can be removed after putting your text content
\usepackage{graphicx}
\usepackage{natbib}
\usepackage{doi}
\usepackage{wrapfig}
\usepackage{amsmath,amsbsy}
\usepackage{amsfonts}
\usepackage{amssymb}
\usepackage{amsthm}
\usepackage{bm}
\usepackage{rotating}
\usepackage{threeparttable}
\usepackage{multirow}
\usepackage{float}
\usepackage{setspace}
\usepackage{booktabs}
\usepackage{threeparttable}
\usepackage{makecell}
\usepackage{graphicx}
\usepackage{hyperref}
\usepackage{authblk}
\usepackage{subfigure}

\usepackage{mathtools}
\usepackage{caption}
\graphicspath{ {./} }
\usepackage{color}

\newcommand{\bY}{ {\bf Y} }
\newcommand{\bX}{ {\bf X} }

\newcommand{\be}{ {\bf e} }

\newcommand{\bA}{ {\bf A} }

\newcommand{\bct}{\begin{center}}
\newcommand{\ect}{\end{center}}

\title{A Random Effects Model-based Method of Moments Estimation of Causal Effect in Mendelian Randomization Studies}

\author{ Wenhao Cao, Saonli Basu\thanks{saonli@umn.edu} \\
	Division of Biostatistics and Health Data Science\\
	University of Minnesota\\
	Minneapolis, Minnesota, USA
}

\renewcommand{\shorttitle}{\textit{arXiv} }

\hypersetup{
pdftitle={A Random Effects Model-based Method of Moments Estimation of Causal Effect in Mendelian Randomization Studies},
pdfsubject={q-bio.NC, q-bio.QM},
pdfauthor={ Wenhao Cao, Saonli Basu},
pdfkeywords={causal inference; instrumental variables; Mendelian randomization; pleiotropy, weak effects; small-scale study},
}

\begin{document}
\maketitle

\begin{abstract}
	Recent advances in genotyping technology have delivered a wealth of genetic data, which is rapidly advancing our understanding of the underlying genetic architecture of complex diseases. Mendelian Randomization (MR) leverages such genetic data to estimate the causal effect of an exposure factor on an outcome from observational studies. In this paper, we utilize genetic correlations to summarize information on a large set of genetic variants associated with the exposure factor. Our proposed approach is a generalization of the MR-inverse variance weighting (IVW) approach where we can accommodate many weak and pleiotropic effects. Our approach quantifies the variation explained by all valid instrumental variables (IVs) instead of estimating the individual effects and thus could accommodate weak IVs. This is particularly useful for performing MR estimation in small studies, or minority populations where the selection of valid IVs is unreliable and thus has a large influence on the MR estimation. Through simulation and real data analysis, we demonstrate that our
approach provides a robust alternative to the existing MR methods. We illustrate the robustness of our proposed approach under the violation of MR assumptions and compare the performance with several existing approaches.
\end{abstract}

% keywords can be removed

\section{Introduction}
Inferring the causal direction between correlated variables is a pervasive issue that cannot be assessed through simple association testing or regression analysis. Mendelian randomization (MR) is a powerful tool for estimating the causal effect of an exposure variable $X$ on the outcome $Y$ by utilizing genetic variants as instrumental variables (IV) $G$ for exposure \citep{emdin2017mendelian,davey2014mendelian}. With the availability of an increasing number of well-powered genome-wide association studies (GWASs) on a growing number of traits, there has been tremendous interest in using genetic variants as IVs. The validity of MR depends on three key assumptions: (a) Relevance: the IVs are associated with $X$;  (b) Independence: there are no unmeasured confounders of the associations between IVs and $Y$; and (c) Exclusion restriction: the IVs affect $Y$ only through the effect on the $X$. MR can be implemented using either individual-level data or summary statistics, employing methods such as the two-stage least squares (2SLS) \citep{baum2003instrumental} or the ratio of coefficients method \citep{didelez2007mendelian}. These methods can be applied using one IV and be extended to accommodate multiple IVs like inverse-variance-weighted (IVW) \citep{ burgess2013mendelian}. However, there are many ways that the key assumptions may be violated.
\begin{figure}[H]
    \centering
    \includegraphics[width=0.6\textwidth]{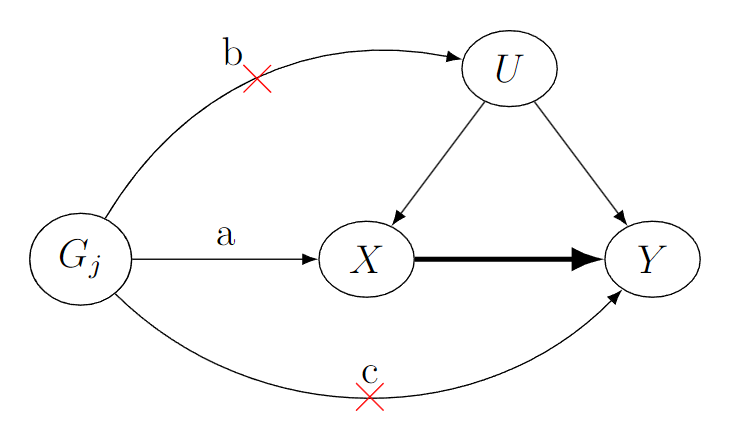}
    \caption{A causal model illustrating the three assumptions on a valid IV}
    \label{Fig1}
\end{figure}

With the proliferation of GWASs, there is a growing tendency to employ a large number of genetic variants as IVs in such investigations \citep{slob2019comparison}. Nevertheless, larger sets of genetic variants are more prone to contain invalid IVs due to horizontal pleiotropy \citep{solovieff2013pleiotropy}, wherein certain IVs may be associated with the outcome variable, thereby violating the exclusion assumption. Another challenge is utilizing IVs with weak effects, which restricts the power to test causal hypotheses and the precision of causal effects. The inclusion of a weak IV can introduce bias, even when the IV satisfies key assumptions and a large sample size is employed \citep{bound1995problems} since a weak IV explains only a small amount of the variation in $X$. An IV weakly associated with the exposure variable yields a small denominator in the ratio estimator, which is commonly used for MR \citep{burgess2017review}, thereby amplifying bias caused by minor violations of the independence assumption and exclusion restriction \citep{hernan2020causal}.

Methods like MR-Egger\citep{bowden2015mendelian}, Simple Median \citep{bowden2016consistent}, Weighted Median\citep{bowden2016consistent}, MR-Lasso \citep{rees2019robust}, have been developed to address the challenge of violating the exclusion restriction. These methods involve the pre-selection of strong IVs to eliminate weak IVs. However, this selection relies on prior knowledge like the F-statistics \citep{bowden2017framework} to select IVs and can potentially exacerbate bias. We call those methods pleiotropy-IV MR methods. To handle the weak IVs without violating the exclusion restriction, methods such as the limited information maximum likelihood (LIML), and continuously updating estimator (CUE), have been proposed \citep{wang2022weak,davies2015many}. These methods, which we call weak-IV MR methods,  aim to mitigate the impact of weak IVs on causal inference.  In the presence of both weak IVs and horizontal pleiotropy, some authors have proposed pleiotropy-weak-IV robust methods that allow for the inclusion of pleiotropic IVs and weak IVs, such as debiased IVW (dIVW) \citep{ye2021debiased}, RAPS\citep{MRRAPS} and genius MR for many weak invalid instruments (GENIUS-MAWII \citep{ye2021genius}). However, these techniques require a large sample size to ensure robustness. With an increasing representation of global populations in a GWAS generally with small sample sizes, there is a strong need to develop approaches that can perform robust MR estimation in a small study or a minority population. It is important to note that with a small sample size, the systematic finite sample bias can be substantial \citep{burgess2011avoiding} in the above-mentioned existing approaches. In this article, we focus on this specific scenario by proposing an approach that performs valid MR estimation in small sample studies by allowing a large number of instrumental variables. We relax the exclusion restriction and allow many weak IVs that belong to the pleiotropy-weak-IV methods category.

 To address these challenges in small-scale studies, we propose a two-stage approach to MR called TS-RE (Two-Stage with Random Effects). Our method shifts the focus from estimating the mean effect sizes of individual IVs to modeling the second-order moment, encompassing the variance and covariance components of the effects of multiple IVs on both the exposure and outcome variables. A distinguishing feature of our approach is the inclusion of many weak IVs, with the simple requirement that the variance of the IVs' effects is non-zero. TS-RE allows all IVs to be weak. The second-order moment estimator in our approach utilizes individual-level data to calculate the genetic correlation matrix (GRM) using the genetic IVs. The IVs included in the analysis are genetic variants that explain a significant proportion of the variance in the exposure variable. By adopting this framework, our proposed approach offers a robust alternative to effectively address the challenges posed by weak and pleiotropic IVs under small sample sizes.

In Section 2, we introduce TS-RE and give theoretical justifications. In Section 3, we conduct simulations to illustrate the superiority of TS-RE in terms of bias reduction, and robustness compared with other methods under a small sample size. We apply TS-RE to investigate the effects of body mass index (BMI) on systolic blood pressure (SBP) using data on the Black British population in the UK Biobank in Section 4. Section 5 concludes the article with a discussion of our findings and the practical implications of TS-RE.

\section{Methods}
\begin{figure}[H]
\centering
\includegraphics[width=0.6\textwidth]{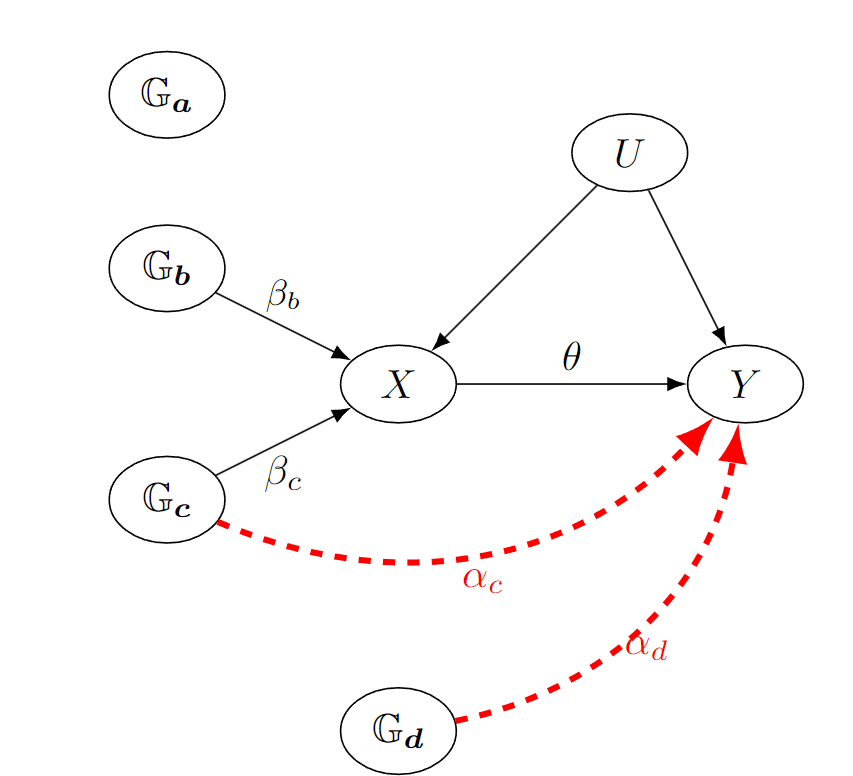}
\caption{A more general Mendelian Randomization model: we are interested in the causal effect $\theta$. Four potential relationships considered: (1) $\mathbb{G}_a$ related to neither $X$ nor $Y$; (2) $\mathbb{G}_b$ with direct effect on $X$ and indirect effect on $Y$; (3) $\mathbb{G}_c$ with direct effects both on $X$ and $Y$; (4) $\mathbb{G}_d$ with direct effect on $Y$ but no relationship with $X$.}
\label{Fig2}
\end{figure}

Suppose the exposure variable $X$, and an outcome $Y$ are causally related, with $X$ affecting $Y$ according to the linear model:
\begin{equation}\label{causal}
Y = \theta X + \epsilon, \quad \epsilon = U + \epsilon',
\end{equation}
The parameter of interest is $\theta$, the causal effect of $X$ on $Y$. The error $\epsilon$ consists of two components: unmeasured confounders $U$ and a residual error term $\epsilon'$. Standard regression estimators like ordinary least squares (OLS) cannot provide consistent unbiased estimates for $\theta$ due to the presence of unmeasured confounders $U$ that are associated with both $X$ and $Y$. To address this issue, MR studies utilize genetic variants, such as Single Nucleotide Polymorphisms (SNPs), as IVs. However, it is important to note that not all selected SNPs are assumed to be valid IVs that satisfy the necessary assumptions aforementioned. As shown in Figure \ref{Fig2}, there are four possible relationships between the IVs and $X, Y$: (1) $\mathbb{G}_{\bm{a}}$ is not related to either $X$ or $Y$; (2) $\mathbb{G}_{\bm{b}}  $ has a direct effect on $X$ and an indirect effect on $Y$; (3) $\mathbb{G}_{\bm{c}}$ has direct effects on both $X$ and $Y$; (4) $\mathbb{G}_{\bm{d}}$ has a direct effect on $Y$ but no relationship with $X$. 

Let's denote $M$ as the total number of IVs, and $M_a, M_b, M_c, M_d$ as the number of IVs in each group. The direct effect of the $k$-th IV ($G_k$) on $X,Y$ is represented by $\beta_k$ and $\alpha_k$, where $k=1,\dots, M$. Therefore, we can express the overall effect of the $k$-th IV on $X$ as $\gamma_{xk}$, and the overall effect on $Y$ as $\gamma_{yk}$:
\begin{equation}
\left\{
\begin{aligned}
    \gamma_{xk} &=0, \gamma_{yk}=0,  \text{if $G_k \in \mathbb{G}_{\bm{a}}$};\\
     \gamma_{xk} &=\beta_{bk}, \gamma_{yk}=\theta\beta_{bk}, \text{if $G_k \in \mathbb{G}_{\bm{b}}$};\\
     \gamma_{xk} &=\beta_{ck}, \gamma_{yk}= \alpha_{ck}+\theta\beta_{ck}, \text{if $G_k \in \mathbb{G}_{\bm{c}}$};\\
     \gamma_{xk} &=0, \gamma_{yk}=\alpha_{dk}, \text{if $G_k \in \mathbb{G}_{\bm{d}}$}.
\end{aligned}
\right.
\end{equation}
where different IVs are assumed independent. Here, the generative model for $X, Y, G$ is
\begin{equation}\label{uniMR}
    \begin{split}
        \bX&=\bm{\gamma_x^TG } +\be_x=\bm{\beta_{b}^T\mathbb{G}_b}+\bm{\beta_{c}^T\mathbb{G}_c} + \be_x,\\
\bY &=\bm{\gamma_y^TG} +\be_y=\theta \bX+\bm{\alpha_{c}^T\mathbb{G}_c }+\bm{\alpha_{d}^T\mathbb{G}_d }+\bm{\epsilon},
    \end{split}
\end{equation}
where $\bm{G}=(\bm{\mathbb{G}_{\bm{a}},\mathbb{G}_{\bm{b}},\mathbb{G}_{\bm{c}},\mathbb{G}_{\bm{d}}})$ is the $n\times M$ matrix of IVs, and $\bm{{\beta}},\bm{{\alpha}}$ are vectors of corresponding coefficients, $\be_x,\be_y$ are errors.

\subsection{Overview of Existing Methods}
First, assuming all IVs are from $\mathbb{G}_{\bm{b}}$, then the 2SLS method can be used if the individual-level data are available:
\begin{equation}
    \hat{\theta}_{2SLS}=\bm{(X^TP_{\mathbb{G}_b}X)^{-1}X^TP_{\mathbb{G}_b}Y}
\end{equation}
where $\bm{P_{\mathbb{G}_b}=\mathbb{G}_b(\mathbb{G}_b^T\mathbb{G}_b)^{-1}\mathbb{G}_b^T}$ is the projection matrix. Let $\bm{G=\mathbb{G}_b}$ in Equation \ref{uniMR}, we can obtain the estimates of coefficients $\bm{\hat{\gamma}_x=(\mathbb{G}_b^T\mathbb{G}_b)^{-1}\mathbb{G}_b^TX}$, $\bm{\hat{\gamma}_y=(\mathbb{G}_b^T\mathbb{G}_b)^{-1}\mathbb{G}_b^TY}$, and variance $se(\bm{\hat{\gamma}_x})^2,se(\bm{\hat{\gamma}_y})^2$ are the diagonal elements of the matrix of $\bm{(\mathbb{G}_b^T\mathbb{G}_b)}^{-1}\sigma^2_{e_x^*}$,$\bm{(\mathbb{G}_b^T\mathbb{G}_b)}^{-1}\sigma^2_{e_y^*}$, where $\sigma_{e_x^*},\sigma_{e_y^*}$ is residual standard error.  If the IVs are perfectly uncorrelated and the effects $\beta_{b1},\dots,\beta_{bM_b}$ are independent, the off-diagonal elements of $\bm{(\mathbb{G}_b^T\mathbb{G}_b)}^{-1}$ are all zero. This means that the 2SLS estimator can be viewed as a weighted average of multiple ratios $\hat{\gamma}{yk}/\hat{\gamma}{xk}$ for $G_k\in \mathbb{G}_{\bm{b}}$. This equivalence allows us to use the IVW method with summary statistics, which is given by:

\begin{equation}\label{IVWregression}
\begin{split}
\hat{\theta}_{IVW}&=\frac{\sum_{k}\hat{\gamma}_{yk}\hat{\gamma}_{xk}se(\hat{\gamma}_{yk})^{-2} }{\sum_{k}(\hat{\gamma}_{xk})^2se(\hat{\gamma}_{yk})^{-2}}\\
&=\bm{(X^TP_{\mathbb{G}_b}X)^{-1}X^TP_{\mathbb{G}_b}Y}
\end{split}
\end{equation}
The IVW estimator will slightly differ from the 2SLS estimator in finite
samples, as the correlation between independent genetic
variants will not exactly equal zero~\citep{burgess2013mendelian}, but the two estimates will be equal asymptotically~\citep{burgess2015mendelian}. The IVW estimator can also be regarded as a regression model as follows:
$$\hat{\gamma
}_{yk}=\theta\hat{\gamma}_{xk}+\epsilon_k,\ \ \epsilon_k\sim N(0, \phi^2_Ise(\hat{\gamma}_{yk})^{2}).$$
If IVs are all from $\mathbb{G}_{\bm{b}}$, then each of the ratios estimates $\hat{\gamma}_{yk}/\hat{\gamma}_{xk}$ will be a consistent estimate of the causal effect $\theta$, thus the 2SLS and IVW (a weighted mean of multiple ratio estimates) will be a consistent estimate of $\theta$. 
$\phi^2_I=1$ is specified under the fixed-effect(FE) IVW without any random intercept if all IVs are valid from $\mathbb{G}_{\bm{b}}$; if some IVs are invalid from $\mathbb{G}_{\bm{c}}$, but the average direct effect of $G$ on $Y$ is zero (referred to as “balanced pleiotropy”), then the model will have a random intercept $\alpha_0 \sim N(0, \tau^2)$ and $\phi^2_I>1$ is assumed under the random-effect IVW. To combat heterogeneity that some IVs are invalid, a random-effects (RE) IVW is used in this paper.   

When IVs are from $\mathbb{G}_{\bm{c}}$ with pleiotropic effect, the ratio $\gamma_{yk}/\gamma_{xk}$ becomes $\theta+\alpha_{ck}/\beta_{ck}$. If the 2SLS and IVW estimators mistakenly include IVs from $\mathbb{G}_{\bm{c}}$ as part of $\mathbb{G}_{\bm{b}}$, they will be biased toward:
\begin{equation}
    \theta+\frac{\sum_{k=1}^{M_c}{\gamma}_{xk}se({\gamma}_{yk})^{-2}\alpha_{ck}}{\sum_{k}({\gamma}_{xk})^2se({\gamma}_{yk})^{-2}}=\theta+Bias(\bm{\alpha_{c},\gamma_c)}.
\end{equation}
This implies that if the exclusion restriction is satisfied all $a_{ck}=0$, 2SLS and IVW estimates are unbiased. However, this will not be universally plausible. To address this issue, the Egger method assumes an average pleiotropic effect for all IVs \citep{burgess2015mendelian}, which assumes that the effects $\beta_k, \alpha_k$ are all random variables. It estimates the average direct effect $\alpha_0$ as part of the analysis, which is assumed to be zero in the IVW method. for Egger.
 Using the same weights in IVW,  the Egger estimator is:
\begin{equation}\label{EggerEstimate}
\begin{split}
    \hat{\gamma
}_{yk}&=\alpha_0+\theta\hat{\gamma}_{xk}+\epsilon_k,\ \ \epsilon_k\sim N(0, \phi^2_Ese(\hat{\gamma}_{yk})^{2})\\
\hat{\theta}_{Egger}&=\frac{Cov_w(\bm{\hat{\gamma}_{y}, \hat{\gamma}_{x})}}{Var_w(\bm{\hat{\gamma}_x})}=\theta+\frac{Cov_w(\bm{\hat{\alpha}_c,\hat{\beta}_c)}}{Var_w(\bm{\hat{\beta})}}
\end{split}
\end{equation}
Due to the potential overdispersion resulting from the pleiotropic effects of IVs, it is recommended to employ a random intercept assuming $\phi^2_E > 1$. In the Egger method, the weighted covariance ($Cov_w$) and weighted variance ($Var_w$) are computed using the inverse-variance weights ($se(\hat{\gamma}_{yk})^{-2}$) and the vector of IV coefficients ($\bm{\beta=(\beta_b,\beta_c)}$). To satisfy the necessary condition for Egger, it is required that the correlation between the effects of IVs on the exposure and the direct effects of IVs on the outcome is zero, denoted as $Cov(\bm{\alpha_c,\beta_c})=0$. This condition, known as the InSIDE (Instrument Strength Independent of Direct Effect) assumption, can be viewed as a weakened version of the exclusion restriction assumption.
Under the InSIDE assumption and as the sample size increases, the weighted covariance $Cov_w(\bm{\hat{\alpha}_c,\hat{\beta}_{c}})$ converges to zero as $n$ tends to infinity, which implies that $Cov_w({\bm{\alpha_c,{\beta_c}}})\xrightarrow[]{n\rightarrow \infty} 0$. Consequently, the Egger estimate becomes a consistent estimate of $\theta$. In Equation \ref{EggerEstimate}, the intercept term $\alpha_0$ represents the average pleiotropic effect of the genetic variants included in the analysis \citep{burgess2013use}.
If $\alpha_0=0$, the estimates obtained by Egger and IVW methods will be the same. However, the standard error of the Egger method will be larger than that of IVW. Instead of using summary statistics for Egger, 
we can expand Equation \ref{EggerEstimate} to show the difference between Egger and 2SLS with individualized data:
\begin{equation}\label{EggerInd}
    \begin{split}
\hat{\theta}_{Egger}&=\frac{\sum_{k}se(\hat{\gamma}_{yk})^{-2}\sum_{k}\hat{\gamma}_{yk}\hat{\gamma}_{xk}se(\hat{\gamma}_{yk})^{-2} -\sum_{k}se(\hat{\gamma}_{yk})^{-2}\hat{\gamma}_{yk}\sum_{k}\hat{\gamma}_{xk}se(\hat{\gamma}_{yk})^{-2}}{\sum_{k}se(\hat{\gamma}_{yk})^{-2}\sum_{k}(\hat{\gamma}_{xk})^2se(\hat{\gamma}_{yk})^{-2}-(\sum_{k}\hat{\gamma}_{xk}se(\hat{\gamma}_{yk})^{-2})^2}\\
        &= \frac{\bm{1^TG^TG1X^TP_GY-X^TG1Y^TG1}}{\bm{1^TG^TG1X^TP_GX-X^TG1X^TG1}}
    \end{split}
\end{equation}
where $\bm{1}$ is a $(M_b+M_c)\times 1$ vector with all elements equal to 1 and $G=(\mathbb{G}_b,\mathbb{G}_c)$.  However, it is unrealistic to assume that the InSIDE assumption always holds all IVs. The limitation of Egger (and related methods) has been discussed  \citep{lin2022practical}, which depends on the orientation of SNPs to get an average pleiotropic effect $\alpha_0$  

Assuming all IVs effects are random variables, we derived the bias term for IVW (2SLS) and Egger if all IVs have a direct effect on $X$ (see Supplemental information Equation \ref{BiasIVW} and Equation \ref{BiasEgger}: 
$Bias_{IVW}=\frac{M_cE(\beta_{ck}\alpha_{ck})}{M_bE(\beta_{bk}^2)+M_cE(\beta_{ck}
        ^2)}$ and $Bias_{Egger}=\frac{M_c}{M} \frac{E(\beta_ck\alpha_ck)-E(\beta_ck)E(\alpha_ck)}{Var(\beta)}$. To get an unbiased estimate, Egger requires the InSIDE assumption that $Var(\alpha_c,\beta_c)=0$, while IVW needs both InSIDE assumption $Var(\alpha_c,\beta_c)=0$ and balanced pleiotropic assumption $E(\alpha_c)=0$.  A selection to avoid IVs with weak effect is required for 2SLS, IVW and Egger, and including IVs from $\mathbb{G}_a$ and $\mathbb{G}_d$ will lead to a large bias to all three methods. 
        
        To address the problem that includes many invalid IVs, the other two commonly used methods Simple Median \citep{bowden2016consistent} and Weighted Median \citep{bowden2016consistent} focus on using the median of $M$ ordered ratio estimator for each IV as the estimate of $\theta$ and allow up to $50\%$ IVs to violate the exclusion restriction. However, the two median estimators are low-powered and sometimes biased when the proportion of invalid IVs is greater than 50$\%$. Furthermore, those biases can be exaggerated under a finite small sample size. Details about these and other existing approaches are listed in Table~\ref{Table1}. 

The methods of 2SLS, IVW, and Egger aim to estimate the causal effect of variable $X$ on variable $Y$ by utilizing the weighted mean of multiple ratios $\gamma_{yk}/\gamma_{xk}$. However, these methods rely on the assumption that the selected IVs have significant and strong effects on the exposure variable $X$. If the selected IVs have weak effects, such as in cases with small sample sizes, these methods may not provide precise estimates of the causal effect. Moreover, including IVs from groups $\bm{\mathbb{G}_a}$ and $\bm{\mathbb{G}_d}$ can lead to misspecification in these existing methods. Table \ref{Table1} compares our proposed TS-RE method with popular existing methods.

\begin{table}[]
    \centering
    \begin{tabular}{cccc}
    \hline
       Method  &  Invalid (c) & Weak IV& Comment\\
       \hline
       TSLS\citep{baum2003instrumental}  & No &No & individual data, $M\leq n$\\
       IVW (FE) \citep{burgess2013mendelian}& No &No & all IVs are valid\\
       IVW (RE) \citep{burgess2013mendelian}& Yes &No & balanced pleiotropy, InSIDE\\
       Egger\citep{burgess2015mendelian}& Yes &No & InSIDE, large SE\\
       Lasso\citep{rees2019robust}& Yes &No & need choose  tuning parameter \\
       Simple Median\citep{bowden2016consistent}& Yes &No & $\leq 50\%$ invalid IVs\\
       Weighted Median\citep{bowden2016consistent}& Yes &No & $\leq 50\%$ invalid IVs\\
       LIML\citep{davies2015many}& No &Yes & individual data, $M\leq n$\\
       CUE\citep{davies2015many}& No & Yes & individual data, $M\leq n$\\
       debiased-IVW\citep{ye2021debiased}&Yes &Yes & balanced pleiotropy, InSIDE\\
       RAPS\citep{MRRAPS}&Yes &Yes & InSIDE\\
       Genus-MAWII\citep{ye2021genius}& Yes & Yes & individual data, $M\leq n$\\
       TS-RE&Yes& Yes&$E(\beta_{ck}\alpha_{ck})=0$\\
       \hline
    \end{tabular}
    \caption{Comparison of different MR methods, including whether IVs violated exclusion restriction and weak IVs are allowed}
    \label{Table1}
\end{table}

\subsection{Our Proposed Approach}
 We introduce a new method that models the variance of multiple IVs' effects instead of estimating individual effect sizes. The TS-RE approach assumes a random effects model for the IVs' effects. It can accommodate a large number of IVs and is less sensitive to the presence of weak IVs. The variance components can be used to estimate the causal effect $\theta$.  The TS-RE method allows for the inclusion of IVs from all four groups in Figure \ref{Fig2}, which makes it more flexible than Egger. However, it requires at least two IVs to come from either $\bm{\mathbb{G}_b}$ or $\bm{\mathbb{G}_c}$ to estimate the variance. In Equation \ref{uniMR}, we assume
\begin{equation}\label{effect}
\begin{split}
    \beta_{bk}&\sim N(\mu_{g_{\bm{b}}},\sigma^2_{g_{\bm{b}}}), k=1,\dots, M_b\\
    \left(
\begin{array}{c}
     \beta_{ck} \\
     \alpha_{ck}
\end{array}
\right)
&\sim
N\left[\bm{\mu_{g_c}},\left(
\begin{array}{cc}
    \sigma_{g_{\bm{c}}^{x}}^2 & \rho_{g_{\bm{c}}}\sigma_{g_{\bm{c}}^{x}}\sigma_{g_{\bm{c}}^{y}}\\\rho_{g_{\bm{c}}}\sigma_{g_{\bm{c}}^{y}}\sigma_{g_{\bm{c}}^{x}}&\sigma_{g_{\bm{c}}^{y}}^2
\end{array}
\right)\right], k=1,\dots, M_c\\
 \alpha_{dk}&\sim N(\mu_{g_d},\sigma^2_{g_{\bm{d}}}), k=1,\dots, M_d
\end{split}
\end{equation}

Here $\mu_{g(\cdot)}$ is the mean effect and $\rho_{\mathbb{G}_{\bm{c}}}$ is the genetic correlation coefficient that quantifies how the InSIDE assumption is violated. $\rho_{\mathbb{G}_{\bm{c}}}=0$ indicates the InSIDE assumption holds. Suppose that (1) residuals $\be_y$, $\be_x$, and $\bm{G}$ are independent of each other; (2) different $G_k$ are independent and the effects of $G_k$ are independent. Then the genetic variances and covariance of $X$ and $Y$ explained by the included IVs are:
\begin{equation}
\begin{split}
   Var_g (\bX) =&\bm{\mathbb{G}_b\mathbb{G}_b^T}E(\beta_{bk}^2)+\bm{\mathbb{G}_c\mathbb{G}_c^T}E(\beta_{ck}^2),\\
    Cov_g (\bX, \bY)=&\theta Var_g(\bX)+\bm{\mathbb{G}_c\mathbb{G}_c^T} E(\alpha_{ck}\beta_{ck})\\
    \xRightarrow[]{E(\alpha_{ck}\beta_{ck})=0}&\theta Var_g(\bX).
\end{split}
\end{equation}
If the assumption that $E(\alpha_{ck}\beta_{ck})=0$ is valid, then the genetic covariance of $X$ and $Y$ is $\theta Var_g(\bX|G)$, which is $\theta$ times the genetic variance of $X$.  Thus, we considered using a second-moment estimator for the genetic variance and covariance of $X, Y$ explained by IVs, instead of means, for the causal estimation where weak IVs are included. After taking the cross-product, the model in Equation \ref{causal} can be written as
\begin{equation}\label{crossP}
\begin{split}
    Y_iX_j&=\theta X_iX_j+ e_{yi}X_j,\ \ \  i,j=1,\dots,n.
\end{split}
\end{equation}
Here the OLS of $\theta$ is biased since $\epsilon_{yi}X_j$ is associated with $X_iX_j$. We involve the second moment to address this problem. In addition, since we can assume different observations are independent, using the covariance of $X_i$ and $X_j$ ($i< j$) to estimate the effect of $G$ on $X$ can allow us to eliminate the variance residual term $X_iX_i$. We perform the following model:
\begin{equation}\label{crossP2}
 X_iX_j =\eta A_{ij}+e^{xx}_{ij},\ \ \ Y_iX_j=\delta A_{ij}+e^{yx}_{ij}.
\end{equation}
where $ i<j;\ i,j=1,\dots,n$, $A_{ij}$ is the $ij$-th element of the GRM, $\bA= \bm{GG^{T}}/M$, and there are $N=\frac{n(n-1)}{2}$ observations. Here the genetic data are standardized and $E(A_{ij}) = 0$. The GRM $\bA$ is used to construct a second-moment estimator.

As a two-stage estimator, the first stage for our TS-RE is to estimate $\hat{\eta}$ and $\hat{\delta}$ in Equation \ref{crossP2}. The second stage is to calculate the ratio $\hat{\theta}_{TS-RE}=\frac{\hat{\eta}}{\hat{\delta}}$.
Similar to the TSLS\citep{baum2003instrumental}, our TS-RE estimator is also equivalent to a  generalized method of moment (GMM) estimator. Denote $\bm{Vec(A), Vec(X\otimes X), Vec(X\otimes Y)}$ to be the $N=n(n-1)/2$ dimensional vectors in Equation \ref{crossP2}. Given certain independence conditions in the  Supplemental information \ref{GMMproff}, we can prove $E[A_{ij}e_{ij}^{yx}))]=0$. Then, the generalized method of moment estimator is given by solving $$g(\hat{\theta})=\frac{1}{N}\bm{Vec(A)}^T(\bm{Vec(X\otimes Y)}-\hat{\theta}\bm{Vec(X\otimes X)})=0,$$ which leads to the following estimator:
\begin{equation}
\begin{split}
    \hat{\theta}_{GMM}
    &=\bm{[Vec(A)^TVec(X\otimes X)]^{-1}Vec(A)^TVec(X\otimes Y)}\\
    &= \widehat{Cov}\left[A_{ij},Y_iX_j\right]/ \widehat{Cov}\left[A_{ij},X_iX_j\right]
\end{split}
\end{equation}

   Through our proof in the Supplemental information \ref{BiasProof}, the bias of our TS-RE estimator $\hat{\theta}_{TS-RE}$ is $\frac{M_cE(\beta_{ck}\alpha_{ck})}{M_bE(\beta_{bk}^2)+M_cE(\beta_{ck}
        ^2)}$. When $E(\beta_{ck}\alpha_{ck})=0$, the estimator 
 will be an unbiased consistent estimator of $\theta$ 
 \begin{equation*}
      \sqrt{N}[\hat{\theta}_{TS-RE}-\theta] \xrightarrow  {\mathcal{D}} N(0, \tau^2)
 \end{equation*}, and
 \begin{equation}\label{aV}
         \tau^2 = \frac{M[M_bE(\beta_{bk}
        ^2)+M_cE(\beta_{ck}
        ^2)+\sigma_{e_x}^2][M_cE(\alpha_{ck}
        ^2)+M_dE(\alpha_{dk}
        ^2)+\sigma_{e_y}^2]}{[M_bE(\beta_{bk}
        ^2)+M_cE(\beta_{ck}
        ^2)]^2}
 \end{equation}
        where $E(\beta^2_{g(\cdot)})=\mu_{g(\cdot)}^2+\sigma_{g(\cdot)}^2$. The key strength of our TS-RE method is that we use SNPs from four groups without selection. IVs from $\mathbb{G}_a, \mathbb{G}_d$ do not contribute to the genetic variance of $X$ and genetic covariance of $X, Y$, thus do not impact the estimation of $\delta$ and $\eta$. The TS-RE method can still have an unbiased estimator even including those invalid IVs.

First, assuming all IVs are from $\mathbb{G}_b$, then the bias term will be $0$ and the asymptotic variance term in Equation \ref{aV} can be written as 
\begin{equation*}
    \tau^2 =(\frac{1}{E(\beta_{bk}
        ^2)}+ \frac{\sigma_{e_x}^2}{M_bE(\beta_{bk}
        ^2)^2})\sigma_{e_y}^2.
\end{equation*}
This indicates for a finite sample that fixing $N$, when all IVs are valid from $\mathbb{G}_b$, stronger effect $\mu_{gb}$, larger variance $\sigma_{gb}^2$, and a larger number of IVs $M_b$  can lead to smaller asymptotic variance. Hence even when the IVs are weak with $mu_{gb} \approx 0$, including a large number of weak IVs that explains a large proportion of the overall exposure variance could give us an efficient estimator. Then, assuming IVs are from $\mathbb{G}_b, \mathbb{G}_c$ and $E(\beta_{bk}^2)=E(\beta_{ck}^2)$,   Equation \ref{aV} can be written as 
\begin{equation*}
    \tau^2 =(\frac{1}{E(\beta_{bk}
        ^2)}+ \frac{\sigma_{e_x}^2}{(M_b+M_c)E(\beta_{bk}
        ^2)^2})M_cE(\alpha_{ck}^2)+\sigma_{e_y}^2.
\end{equation*}
Including a large number of IVs with direct effects on exposure variable $M_b+M_c$ can control the variance of the TS-RE estimator even if the IVs' effects are weak. Including IVs with directional pleiotropic effects $\mu_{\mathbb{G}_c^y}\neq 0$ leads to a larger $\tau^2$. Among IVs with direct effects on $X$ from $\mathbb{G}_b, \mathbb{G}_c$, the variance will increase if the proportion of IVs from $\mathbb{G}_c$ increases. When IVs from $\mathbb{G}_c, \mathbb{G}_d$ is also included, a higher proportion of null IVs from groups $\mathbb{G}_a, \mathbb{G}_d$ will lead to a larger SE. 
 
\section{Simulations}
\subsection{Simulation Setting}
We performed simulations to study the performance of TS-RE in comparison with five pleiotropy-IV MR approaches based on summary statistics: Simple Median, Weighted Median, IVW, Egger, and Lasso. We also included two pleiotropy-weak-IV methods, dIVW and RAPS. We did not include TSLS and the weak-IV methods using individual-level data since they require the sample size to be larger than the number of IVs.

We considered $M=M_a+M_b+M_c+M_d$ variants from four groups in Figure \ref{Fig2}. Note that variants in $\mathbb{G}_{\bm{b}}$ will be valid IVs. For the summary statistics-based approaches, we performed a simple linear regression for each IV to get the summarized statistics. We used all IVs for the TS-RE method and the top 20 significant (based on p-value) IVs for other methods. Considering top significant variants will generally eliminate the weak IVs and invalid IVs in $\mathbb{G}_{\bm{a}}$ and $\mathbb{G}_{\bm{d}}$ which is required for other methods.  For each simulation setup, we generated 100 datasets and estimated the causal effects. We reported the bias and the standard error (SE) of $\theta$ from these 100 simulations.

We followed the following procedure for each simulation. First, the minor allele frequency (MAF) $f_k$ of each SNP $k$ ($k=1,\dots, M$) was independently generated from a uniform distribution $U(0.2, 0.3)$ and the corresponding genotypes were simulated from a binomial distribution $Bin(2, f_k)$. Second, we generate the effects of IVs from Equation \ref{effect}. We used different $\mu_{(\cdot).g},\sigma_{(\cdot)g}^2$ to generate IVs effects.  Assuming the total variance for $X$ and $Y$ are $\sigma_X^2$ and $\sigma_Y^2$, where $\sigma_X^2=\sigma_{G_x}^2+\sigma_{e_x}^2,\sigma_Y^2=\sigma_{G_y}^2+\sigma_{e_y}^2$, where $ \sigma_{G_x}^2, \sigma_{G_y}^2$ are the total genetic variances of included SNPs. If only IVs from $\mathbb{G}_b$ were included, then $\sigma^2_{G_x}=M_b\sigma_{gb}^2$, $\sigma^2_{G_y}=\theta^2M_b\sigma_{gb}^2$, and the residual variance component are $\sigma_{e_y}^2,\sigma_{e_x}^2$. The proportion of total variance explained by included IVs is defined as conditional heritability (Her) $Her_X=\sigma_{G_x}^2/\sigma_X^2$. $X$ and $Y$ were generated from Equation \ref{uniMR}. The primary objective was to estimate the causal effect $\theta$. 

We compared our TS-RE method with existing methods given a small sample size of $n=1000$ under three scenarios: (1) only $M_b$ IVs from $\mathbb{G}_b$; (2) a mixture of IVs from $\mathbb{G}_b$ and $\mathbb{G}_c$; (3) a mixture of IVs from all four groups. For IVs from $\mathbb{G}_b, \mathbb{G}_c$ having direct effects on $X$, we varied the proportion of IVs with weak effects on $X$.  For IVs with pleiotropic effects from $\mathbb{G}_c$, the effects were generated from four sub-scenarios: (a) Balanced pleiotropy ($ \mu_{\alpha_c}=0$), InSIDE assumption satisfied($\rho_{g_c}=0$);
(b) Directional pleiotropy ($ \mu_{\alpha_c}=0.1$), InSIDE assumption satisfied($\rho_{g_c}=0$);
(c) Balanced pleiotropy ($ \mu_{\alpha_c}=0$), InSIDE assumption violated ($\rho_{g_c}=0.6$);
(d) Directional pleiotropy ($ \mu_{\alpha_c}=0.1$), InSIDE assumption violated ($\rho_{g_c}=0.6$).

\subsection{IVs from $\mathbb{G}_b$: number of IVs and Her}

\begin{figure}[H]
\centering

\includegraphics[width=1\textwidth]{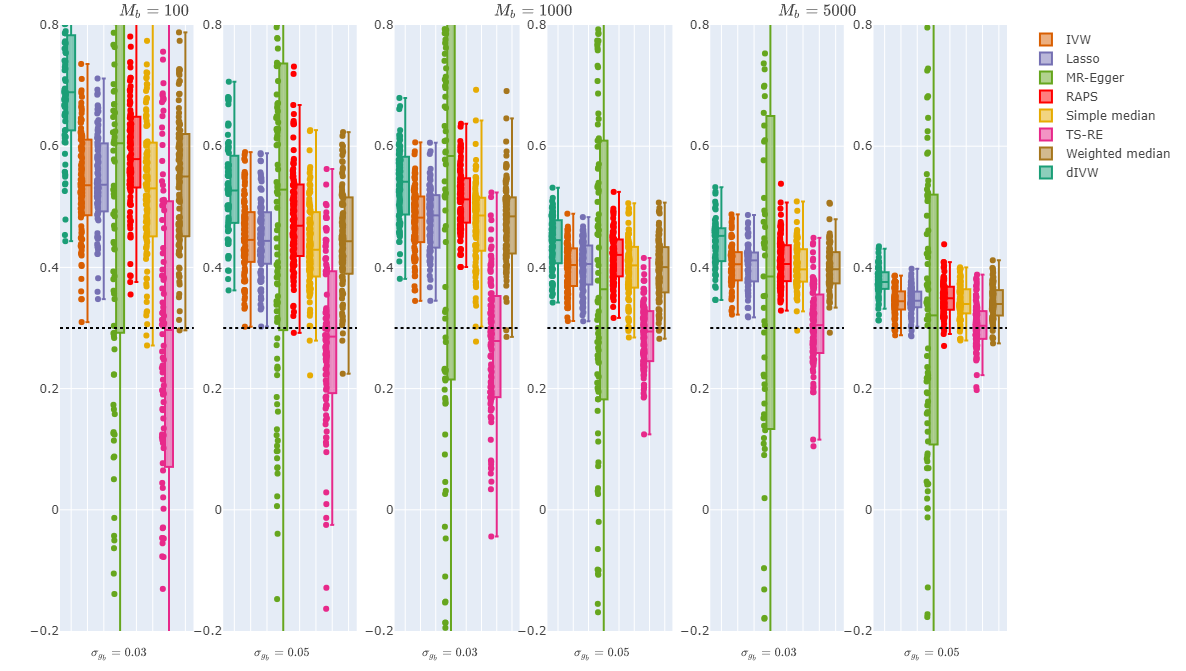}
\caption{ Empirical distributions of the estimates of the causal effect $\theta=0.3$ by the methods with different numbers of IVs and different genetic variances. TS-RE used all IVs while other MR methods used the selected top 20 most significant IVs.}
\label{Fig3}
\end{figure}

In this simulation study, we set parameters: $M_a=0, M_c=0, M_d=0$ and the effects of IVs with $\beta_{bk}\sim N(0, \sigma_{gb}^2)$ for $k=1,\dots, M_b$, explored different values of $\sigma_{gb}= 0.03, 0.05$ and numbers of IVs $M_b=100,1000,5000$. The causal effect $\theta=0.3$, and the residual variance $\sigma_{e_x}^2=2$. Figure~\ref{Fig3} presents results for TS-RE using all IVs and other MR methods employing the top 20 significant IVs.
TS-RE remained unbiased, albeit with larger standard errors, unlike other methods.  While other MR methods exhibited larger biases, our method stood out by consistently providing unbiased estimates even with weak IVs. Our TE-RE method exhibited decreased bias with larger Her, which was achievable by higher $\sigma_{g_b}^2$ or larger $M_b$. Increasing $\sigma_{g_b}^2$ also resulted in lower biases for other methods, likely because larger effect values $\beta$ could be more easily generated, reducing the impact of weak IVs on the results. These findings underscore that a larger Her does not always guarantee better performance for other MR methods, as previously suggested by Freeman et al. \citep{freeman2013power}. Increasing Her through the inclusion of numerous weak IVs did not consistently improve the performance of other MR methods. Instead, the efficiency of the TS-RE method appears to benefit more when the variants collectively explain a larger proportion of the variance in the exposure, indicating its potential advantage under such circumstances.

\begin{figure}[H]
\centering

\includegraphics[width=1\textwidth]{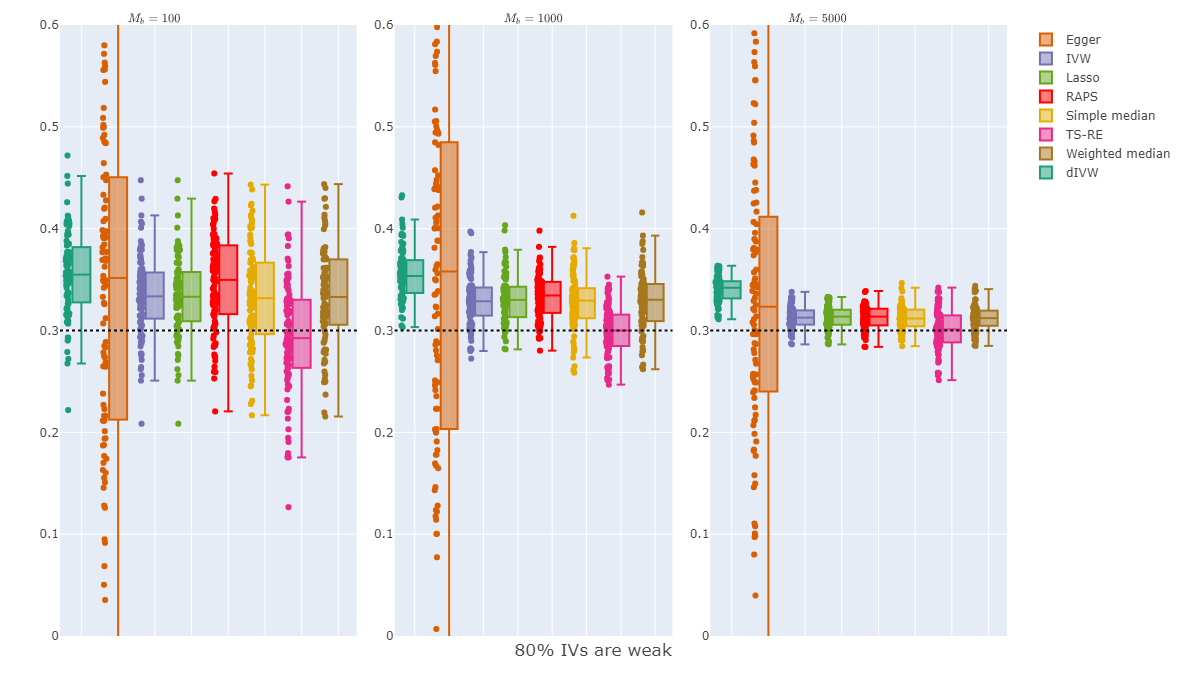}
\caption{ Empirical distributions of the estimates of the causal effect $\theta=0.3$ by the methods with different numbers of IVs and $80\%$ IVs are weak. TS-RE used all IVs while other MR methods used the selected top 20 most significant IVs.} 
\label{Fig4}
\end{figure}

Considering that other methods require some IVs with strong effects, we introduced a scenario where $20\%$ of the IVs had strong effects generated from $N(0.2, 0.05^2)$, while the remaining IVs had weak effects generated from $N(0, 0.05^2)$. We explored different $M_b=100,1000,5000$, while setting $\theta=0.3$ and $\sigma_{e_x}^2=2$. The results are depicted in Figure ~\ref{Fig4} for scenarios where all IVs were weak and when $80\%$ of the IVs were weak. The inclusion of IVs with strong effects led to a smaller bias for the other methods, although some bias remained. Notably, the TS-RE method exhibited the least bias across all configurations and approached an unbiased estimate when the number of IVs exceeded 500.

Incorporating  IVs with strong effects also contributed to a reduction in the SE of the TS-RE method, consistent with our proof in Equation \ref{aV}. However, the inclusion of IVs with strong effects did not notably enhance the performance of the other methods due to the limited sample size. It was observed that the dIVW and RAPS methods were sensitive to the small sample size, displaying significant bias when all IVs were weak. For dIVW, this might be attributed to its reliance on a large sample size to yield a precise estimate for $\hat{\gamma}_x$ and $se(\hat{\gamma}_x)$, which are crucial for adjusting the bias introduced by IVW. For RAPS, the profile likelihood also required a large sample size. In supplemental Table S3, we show that TS-RE has a similar performance with other methods with large sample sizes.

\subsection{IVs from $\mathbb{G}_c$: pleiotropic effect and InSIDE assumption}
Based on the results of the previous simulation, we excluded the dIVW and RAPS methods due to their poor performance under a small sample size. We omitted the Lasso method as its reliability depended on additional information about how to select the tuning parameter. We deliberately chose a small number of IVs ($M_b=M_c=100$) and set genetic variances as $\sigma_{g_b^x}=\sigma_{g_c^x}=\sigma_{g_c^y}=0.03$, along with residual variances of $\sigma_{e_x}^2=\sigma_{e_y}^2=2$. The causal effect was $\theta=0.3$. We checked the scenarios that all IVs were weak and $80\%$ were weak.

Table~\ref{Table2} illustrates the impact of pleiotropy and the validity of the InSIDE assumption. Here are the findings: (1) when balanced pleiotropy and the InSIDE assumption were met, TS-RE consistently yielded nearly unbiased estimates, outperforming the other MR methods when all IVs were used. (2) When directional pleiotropy was introduced while satisfying the InSIDE assumption, the performance of TS-RE declined, particularly when $20\%$ of the $M_c$ IVs had strong effects. This violated the unbiasedness requirement of TS-RE, which assumes that $E(\beta_c\alpha_c)=0$. However, when all $M_c$ IVs were weak, TS-RE's bias remained smaller than other MR methods. It's worth noting that, due to the small sample size, Egger could not provide unbiased estimates, even though it was designed to handle pleiotropy. (3) When the InSIDE assumption was violated ($Var(\beta_c,\alpha_c)\neq 0$), thus $E(\beta_c\alpha_c)\neq 0$, which is a key assumption for TS-RE. None of the methods, including TS-RE, were able to provide unbiased estimates under this condition. In summary, the performance of TS-RE was generally robust when the InSIDE assumption was satisfied and pleiotropy was balanced. It was sensitive to directional pleiotropy when strong IV effects were introduced. We also investigated the impact of proportions $M_b:M_c$ in Supplemental Table \ref{S4}.

\begin{table}
\small
    \centering
    \begin{tabular}{ccccccc}
    \hline
        & $p_w$ & SM & WM & IVW & Egger  & TS-RE\\
       \hline
  Balanced pleiotropy,   & 0.8 &   0.32 (0.06) & 0.33 (0.07) & 0.33 (0.05) & 0.40 (0.30) &  0.29 (0.05)\\
      \cline{2-7}
   InSIDE satisfied  
    &    1  & 0.43 (0.11) & 0.44 (0.11) & 0.43 (0.09) & 0.42 (0.61) & 0.29 (0.16)\\
     \hline
   Directional pleiotropy,   &    0.8 &  0.70 (0.07) & 0.70 (0.07) & 0.70 (0.06) & 0.63 (0.34)  & 0.68 (0.07)\\
      \cline{2-7}
   InSIDE satisfied 
       & 1     &  0.42 (0.16) & 0.41 (0.17) & 0.42 (0.15)  & 0.32 (1.00)  & 0.27 (0.24)\\
     \hline
      
  Balanced pleiotropy,  &  0.8 & 0.76 (0.12) & 0.78 (0.12) & 0.77 (0.11) & 0.82 (0.48)  & 0.73 (0.09) \\
      \cline{2-7}
    InSIDE violated  
      & 1     & 0.81 (0.13) & 0.80 (0.13) & 0.79 (0.12) & 0.74 (0.49) & 0.78 (0.10)\\
     \hline
       
     Directional pleiotropy,  & 0.8  &   0.80 (0.12) &0.83 (0.13)  &0.84 (0.11)  & 0.93 (0.51) & 0.79 (0.09)\\
      \cline{2-7}
    InSIDE violated   
     &  1   & 0.89 (0.14) & 0.89 (0.14) & 0.89 (0.12) & 0.87 (0.67)  & 0.87 (0.10)\\
     \hline
    \end{tabular}
    \caption{Mean and SE of different methods for $M_b=100,M_c=100$ IVs: under different pleiotropic effect and InSIDE assumption conditions. The directional effect is 0.1 and the genetic correlation is 0.6 when the InSIDE assumption is invalid. The strong effect for some valid $M_b$ IVs is 0.2.}
    \label{Table2}
\end{table}

\subsection{Null IVs from group $\mathbb{G}_a$ and $\mathbb{G}_d$}
Here we included all four groups of IVs.  Weak IV effects $\beta_{bk}, \beta_{ck}$on $X$ were generated $N(0, 0.03^2)$, while strong effects were generated $N(0.2, 0.03^2)$. Effects $\alpha_{ck}, \alpha_{dk}$ on $Y$ were generated $N(0, 0.03^2)$. We set $\theta=0.3$, and $\sigma_{e_x}^2=\sigma_{e_y}^2=2$. The IVs from group $\mathbb{G_c}$ satisfied the balanced pleiotropic assumption and the InSIDE assumption. The number of IVs from each group was set to 100, 200, and 500, while the total number of IVs was 400, 800, and 2000. 
\begin{table}[]
    \centering
    \begin{tabular}{cccccccc}
    \hline
 IVs in each group  & method &\multicolumn{3}{c}{All IVs}&\multicolumn{3}{c}{Top20 IVs}\\
 \cline{3-8}
& &Bias&SE&MSE&Bias&SE&MSE\\
 \hline
\multirow{5}{*}{100}&SM&0.2&0.09&0.05&0.15&0.1&0.03\\
&WM&0.17&0.07&0.03&0.15&0.11&0.03\\
&IVW&0.18&0.06&0.04&0.15&0.08&0.03\\
&Egger&0.16&0.07&0.03&0.13&0.61&0.39\\
&TS-RE&-0.03&0.2&0.04&0.13&0.11&0.03\\
\hline
\multirow{5}{*}{200}&SM&0.17&0.06&0.03&0.14&0.1&0.03\\
&WM&0.16&0.06&0.03&0.14&0.1&0.03\\
&IVW&0.17&0.05&0.03&0.14&0.08&0.03\\
&Egger&0.15&0.07&0.03&0.07&0.73&0.54\\
&TS-RE&0.003&0.15&0.02&0.13&0.1&0.03\\
\hline
\multirow{5}{*}{500}&SM&0.13&0.05&0.02&0.1&0.09&0.02\\
&WM&0.12&0.05&0.02&0.1&0.09&0.02\\
&IVW&0.12&0.05&0.02&0.11&0.08&0.02\\
&Egger&0.12&0.06&0.02&-0.05&0.71&0.5\\
&TS-RE&0.02&0.14&0.02&0.12&0.1&0.02\\
\hline
    \end{tabular}
    \caption{Simulation results for the mixture of IVs from four groups. The number of IVs from each group is equal set to be $100,200,500$, while the total number of IVs is $400,800,2000$. The IVs with the direct effect on exposure from $\mathbb{G}_b$ and $\mathbb{G}_c$ have an effect from a normal distribution $N(0,0.03^2)$. IVs from $\mathbb{G}_c$ have balanced pleiotropy and the InSIDE assumption is valid.}
    \label{Table3}
\end{table}

TS-RE consistently provided more unbiased estimates than other methods, while the SE was larger and the Mean Squared Error (MSE) was similar compared to other methods in Table \ref{Table3}. We increased the number of IVs in $\mathbb{G}_a$ from 1000 to 50000 while maintaining the number of IVs in the other groups at 1000. The results revealed that including too many $\mathbb{G}_a$ null IVs resulted in increased bias for TS-RE. There was a significant increase in SE as the number of null IVs increased, particularly considering that the sample size was much smaller than the number of IVs. This observation aligns with the theoretical result that $\tau^2\propto \frac{M}{M_b+M_c}$.

\begin{table}
    \centering
    \begin{tabular}{ccc}
    \hline
     $M_a$ & Estimate & SE\\
     \hline
    1000& 0.3047400& 0.09359366\\
   2000 &0.2962185& 0.12384585\\
   5000& 0.2859627& 0.26879603\\
  10000& 0.2915771& 0.21386340\\
  20000& 0.2859134 &0.38467096\\
   50000 &0.1124109 &2.40895732\\
   \hline
    \end{tabular}
    \caption{Change the number of null IVs from $\mathbb{G}_a$ from 1000 to 50000 where the number of each other three groups is fixed to be 1000.}
    \label{Table4}
\end{table}

\section{Real data analysis}
We estimated the causal effect of BMI on SBP for black British individuals from the UK-biobank dataset, with a total sample size of 3396. Following quality control procedures and linkage disequilibrium pruning, we retained 151442 SNPs with an allele frequency of 0.05, HWE of 0.000001, MAF of 0.01, LD window of 1000, the step of 50, and $r^2$ of 0.1. Our sample size was 2802 after removing the related individuals with GRM cutoff $\geq 0.05$. We applied our TS-RE method and four other MR methods to estimate the causal effect of BMI on SBP.

Initially, TS-RE using all SNPs without selection yielded an imprecise result with a large standard error: $\theta=0.31, SE= 0.41$. Here $\theta=0.31$ means a one $kg/m^2$unit increase in BMI increases SBP by $0.31 mmHg$. This is consistent with previous studies using MR methods with large samples\citep{lyall2017association}. Subsequently, we selected the top 20 significant SNPs for BMI, resulting in nonsensical results for all methods (in Table \ref{Table5}). To avoid including an extremely large number of null IVs in $\mathbb{G}_a$ and $\mathbb{G}_d$, we selected 7580 SNPs with a $p_{value}<0.005$, which reduced the SE of the TS-RE to $0.01$ and yielded an estimate of $\theta=0.33$. In Table \ref{Table6} We attempted to use 56 SNPs as IVs for this small study, where those SNPs were identified by the previous study focused on the white population\citep{locke2015genetic}. Other MR methods failed to estimate the causal effect (large SE, $p-value>0.05$) due to the small sample size.

\begin{table}[!ht]
    \centering
    \begin{tabular}{ccccc}
    \hline
 TS-RE&SM &WM &IVW  & Egger\\
 \hline
 0.31 (0.41) & 2.42 (0.25)&2.42 (0.25)& 2.27 (0.17)& 2.63 (0.42)\\
\hline
    \end{tabular}
    \caption{Causal effect of BMI on SBP for independent black British: TS-RE used all SNPs and the other MR methods used the selected top 20 significant SNPs}
    \label{Table5}
\end{table}

\begin{table}[!ht]
    \centering
    \begin{tabular}{cccc}
    \hline
 SM &WM &IVW& Egger \\
 \hline
  -0.195 (0.871)&0.748 (0.819)& -0.093 (0.639)& 0.949 (0.892)\\
\hline
    \end{tabular}
    \caption{Causal effect of BMI on SBP for independent black British with selected 56 SNPs based on an external study.}
    \label{Table6}
\end{table}

\section{Discussion}
The proposed TS-RE estimator offers a promising solution to address the challenges faced in MR analyses, particularly in small-scale studies. These challenges often include issues related to weak IVs and pleiotropic effects, which traditionally require a large sample size to effectively detect and estimate. In contrast to existing MR approaches that primarily rely on first-order moments for estimating causal effects, our TS-RE method leverages second-order moments to estimate the causal effect of the exposure variable on the outcome variable. By utilizing a substantial number of genetic variants, TS-RE enables the estimation of genetic variances for both the exposure and the outcome variables, ultimately providing an unbiased estimate of the causal effect. Remarkably, this can be achieved even with a small sample size or when dealing with a minority population.

In practice, GWAS often generate data with a large number of SNPs through advanced DNA sequencing techniques, regardless of the sample size. Our novel TS-RE method capitalizes on this wealth of genetic information, allowing for the application of MR to estimate causal effects in small studies or within specific sub-populations of interest within a larger dataset. Furthermore, real data analysis has demonstrated the limitations of conventional MR methods, even when utilizing significant SNPs identified by large external studies. This is particularly evident when attempting to rectify the issue of small sample sizes, primarily due to the heterogeneity between the smaller, specific population of interest and the larger external population. TS-RE offers a robust alternative in such scenarios, where commonly used MR methods relying on large sample sizes may not be feasible or effective.

The TS-RE method offers several key advantages supported by both theoretical insights and empirical simulations. These advantages make it a valuable tool for MR analyses. First, TS-RE does not require strict selection criteria for IVs, such as a specific p-value threshold (e.g., p-value $\leq 5 \times 10^{-8}$). It can handle weak IVs without significantly amplifying biases resulting from the violation of the exclusion restriction. This means that even including some null IVs is acceptable for TS-RE, providing greater flexibility in IV selection. Second, unlike many other MR methods that require a very large sample size for consistency, TS-RE achieves consistency even with a small sample size by incorporating a large number of IVs. This sets it apart from first-moment-based MR methods that depend on large sample sizes for reliable estimates. TS-RE excels in small studies and performs comparably to other MR methods in larger studies. Third, theoretical analysis shows that TS-RE is equivalent to the IVW method when all IVs have a direct effect on the exposure variable. After obtaining the Genetic Relationship Matrix (GRM), the application of TS-RE is as straightforward as the IVW estimator, simplifying the estimation process. Although the TS-RE prefers a large number IVs to estimate the genetic variance and covariance, as we showed in the simulation study, the performance of our TS-RE was not worse than other methods using the selected top 20 significant IVs.

We acknowledge the certain limitations of TS-RE. First, TS-RE tends to have larger standard errors compared to other MR methods, primarily because it employs a second-moment estimator, which introduces more uncertainty. Despite this, TS-RE significantly reduces bias and achieves similar MSE compared to other MR methods. This makes it a valuable alternative for obtaining more unbiased estimates in practical settings. In addition, while TS-RE relaxes the strict exclusion restriction, it still assumes that the expectation of the product of pleiotropic effects is zero, as specified by conditions like InSIDE. This assumption may have a limited biological basis, as it restricts unknown pleiotropic effects of SNPs. However, when a large number of weak IVs are included, the average of the product approximates zero. Future developments of TS-RE may explore ways to further relax both the independence and exclusion restrictions, potentially aligning with approaches like MR-Genius\citep{tchetgen2021genius}.

In summary, TS-RE presents a robust and flexible approach to MR analyses, particularly suited for small studies, weak IVs, and scenarios where other MR methods may struggle. Its theoretical foundation and performance in simulations demonstrate its potential as a valuable tool for causal inference in a variety of research contexts. Future work may focus on refining and extending TS-RE to address additional challenges and broaden its applicability in MR analyses. The proposed framework for small studies can also be extended to integrate information across multiple studies \citep{bowden2019meta, burgess2011avoiding}. Using a meta-analysis to combine the estimates of genotype-phenotype association from different studies can give more precise estimates of the IVs effect. Similarly, the current analysis of a small sub-population can be extended to multiple sub-populations to investigate the causal effect in the presence of population sub-structure \citep{lin2022estimating}. 

\bibliographystyle{unsrtnat}
\bibliography{references}

\begin{thebibliography}{29}
\providecommand{\natexlab}[1]{#1}
\providecommand{\url}[1]{\texttt{#1}}
\expandafter\ifx\csname urlstyle\endcsname\relax
  \providecommand{\doi}[1]{doi: #1}\else
  \providecommand{\doi}{doi: \begingroup \urlstyle{rm}\Url}\fi

\bibitem[Emdin et~al.(2017)Emdin, Khera, and Kathiresan]{emdin2017mendelian}
Connor~A Emdin, Amit~V Khera, and Sekar Kathiresan.
\newblock Mendelian randomization.
\newblock \emph{Jama}, 318\penalty0 (19):\penalty0 1925--1926, 2017.

\bibitem[Davey~Smith and Hemani(2014)]{davey2014mendelian}
George Davey~Smith and Gibran Hemani.
\newblock Mendelian randomization: genetic anchors for causal inference in epidemiological studies.
\newblock \emph{Human molecular genetics}, 23\penalty0 (R1):\penalty0 R89--R98, 2014.

\bibitem[Baum et~al.(2003)Baum, Schaffer, and Stillman]{baum2003instrumental}
Christopher~F Baum, Mark~E Schaffer, and Steven Stillman.
\newblock Instrumental variables and gmm: Estimation and testing.
\newblock \emph{The Stata Journal}, 3\penalty0 (1):\penalty0 1--31, 2003.

\bibitem[Didelez and Sheehan(2007)]{didelez2007mendelian}
Vanessa Didelez and Nuala Sheehan.
\newblock Mendelian randomization as an instrumental variable approach to causal inference.
\newblock \emph{Statistical methods in medical research}, 16\penalty0 (4):\penalty0 309--330, 2007.

\bibitem[Burgess et~al.(2013)Burgess, Butterworth, and Thompson]{burgess2013mendelian}
Stephen Burgess, Adam Butterworth, and Simon~G Thompson.
\newblock Mendelian randomization analysis with multiple genetic variants using summarized data.
\newblock \emph{Genetic epidemiology}, 37\penalty0 (7):\penalty0 658--665, 2013.

\bibitem[Slob and Burgess(2019)]{slob2019comparison}
Eric~AW Slob and Stephen Burgess.
\newblock A comparison of robust mendelian randomization methods using summary data.
\newblock \emph{BioRxiv}, page 577940, 2019.

\bibitem[Solovieff et~al.(2013)Solovieff, Cotsapas, Lee, Purcell, and Smoller]{solovieff2013pleiotropy}
Nadia Solovieff, Chris Cotsapas, Phil~H Lee, Shaun~M Purcell, and Jordan~W Smoller.
\newblock Pleiotropy in complex traits: challenges and strategies.
\newblock \emph{Nature Reviews Genetics}, 14\penalty0 (7):\penalty0 483--495, 2013.

\bibitem[Bound et~al.(1995)Bound, Jaeger, and Baker]{bound1995problems}
John Bound, David~A Jaeger, and Regina~M Baker.
\newblock Problems with instrumental variables estimation when the correlation between the instruments and the endogenous explanatory variable is weak.
\newblock \emph{Journal of the American statistical association}, 90\penalty0 (430):\penalty0 443--450, 1995.

\bibitem[Burgess et~al.(2017)Burgess, Small, and Thompson]{burgess2017review}
Stephen Burgess, Dylan~S Small, and Simon~G Thompson.
\newblock A review of instrumental variable estimators for mendelian randomization.
\newblock \emph{Statistical methods in medical research}, 26\penalty0 (5):\penalty0 2333--2355, 2017.

\bibitem[Hern{\'a}n and Robins(2020)]{hernan2020causal}
Miguel~A Hern{\'a}n and James~M Robins.
\newblock \emph{Causal Inference: What If}.
\newblock CRC Boca Raton, FL, 2020.

\bibitem[Bowden et~al.(2015)Bowden, Davey~Smith, and Burgess]{bowden2015mendelian}
Jack Bowden, George Davey~Smith, and Stephen Burgess.
\newblock Mendelian randomization with invalid instruments: effect estimation and bias detection through egger regression.
\newblock \emph{International journal of epidemiology}, 44\penalty0 (2):\penalty0 512--525, 2015.

\bibitem[Bowden et~al.(2016)Bowden, Davey~Smith, Haycock, and Burgess]{bowden2016consistent}
Jack Bowden, George Davey~Smith, Philip~C Haycock, and Stephen Burgess.
\newblock Consistent estimation in mendelian randomization with some invalid instruments using a weighted median estimator.
\newblock \emph{Genetic epidemiology}, 40\penalty0 (4):\penalty0 304--314, 2016.

\bibitem[Rees et~al.(2019)Rees, Wood, Dudbridge, and Burgess]{rees2019robust}
Jessica~MB Rees, Angela~M Wood, Frank Dudbridge, and Stephen Burgess.
\newblock Robust methods in mendelian randomization via penalization of heterogeneous causal estimates.
\newblock \emph{PloS one}, 14\penalty0 (9):\penalty0 e0222362, 2019.

\bibitem[Bowden et~al.(2017)Bowden, Del Greco~M, Minelli, Davey~Smith, Sheehan, and Thompson]{bowden2017framework}
Jack Bowden, Fabiola Del Greco~M, Cosetta Minelli, George Davey~Smith, Nuala Sheehan, and John Thompson.
\newblock A framework for the investigation of pleiotropy in two-sample summary data mendelian randomization.
\newblock \emph{Statistics in medicine}, 36\penalty0 (11):\penalty0 1783--1802, 2017.

\bibitem[Wang and Kang(2022)]{wang2022weak}
Sheng Wang and Hyunseung Kang.
\newblock Weak-instrument robust tests in two-sample summary-data mendelian randomization.
\newblock \emph{Biometrics}, 78\penalty0 (4):\penalty0 1699--1713, 2022.

\bibitem[Davies et~al.(2015)Davies, von Hinke Kessler~Scholder, Farbmacher, Burgess, Windmeijer, and Smith]{davies2015many}
Neil~M Davies, Stephanie von Hinke Kessler~Scholder, Helmut Farbmacher, Stephen Burgess, Frank Windmeijer, and George~Davey Smith.
\newblock The many weak instruments problem and mendelian randomization.
\newblock \emph{Statistics in Medicine}, 34\penalty0 (3):\penalty0 454--468, 2015.

\bibitem[Ye et~al.(2021{\natexlab{a}})Ye, Shao, and Kang]{ye2021debiased}
Ting Ye, Jun Shao, and Hyunseung Kang.
\newblock Debiased inverse-variance weighted estimator in two-sample summary-data mendelian randomization.
\newblock \emph{The Annals of statistics}, 49\penalty0 (4):\penalty0 2079--2100, 2021{\natexlab{a}}.

\bibitem[Zhao et~al.(2020)Zhao, Wang, Hemani, Bowden, and Small]{MRRAPS}
Qingyuan Zhao, Jingshu Wang, Gibran Hemani, Jack Bowden, and Dylan~S. Small.
\newblock {Statistical inference in two-sample summary-data Mendelian randomization using robust adjusted profile score}.
\newblock \emph{The Annals of Statistics}, 48\penalty0 (3):\penalty0 1742 -- 1769, 2020.
\newblock \doi{10.1214/19-AOS1866}.
\newblock URL \url{https://doi.org/10.1214/19-AOS1866}.

\bibitem[Ye et~al.(2021{\natexlab{b}})Ye, Liu, Sun, and Tchetgen]{ye2021genius}
Ting Ye, Zhonghua Liu, Baoluo Sun, and Eric~Tchetgen Tchetgen.
\newblock Genius-mawii: For robust mendelian randomization with many weak invalid instruments.
\newblock \emph{arXiv preprint arXiv:2107.06238}, 2021{\natexlab{b}}.

\bibitem[Burgess et~al.(2011)Burgess, Thompson, and Collaboration]{burgess2011avoiding}
Stephen Burgess, Simon~G Thompson, and CRP CHD~Genetics Collaboration.
\newblock Avoiding bias from weak instruments in mendelian randomization studies.
\newblock \emph{International journal of epidemiology}, 40\penalty0 (3):\penalty0 755--764, 2011.

\bibitem[Burgess and Thompson(2015)]{burgess2015mendelian}
Stephen Burgess and Simon~G Thompson.
\newblock \emph{Mendelian randomization: methods for using genetic variants in causal estimation}.
\newblock CRC Press, 2015.

\bibitem[Burgess and Thompson(2013)]{burgess2013use}
Stephen Burgess and Simon~G Thompson.
\newblock Use of allele scores as instrumental variables for mendelian randomization.
\newblock \emph{International journal of epidemiology}, 42\penalty0 (4):\penalty0 1134--1144, 2013.

\bibitem[Lin et~al.(2022{\natexlab{a}})Lin, Pan, and Pan]{lin2022practical}
Zhaotong Lin, Isaac Pan, and Wei Pan.
\newblock A practical problem with egger regression in mendelian randomization.
\newblock \emph{PLoS genetics}, 18\penalty0 (5):\penalty0 e1010166, 2022{\natexlab{a}}.

\bibitem[Freeman et~al.(2013)Freeman, Cowling, and Schooling]{freeman2013power}
Guy Freeman, Benjamin~J Cowling, and C~Mary Schooling.
\newblock Power and sample size calculations for mendelian randomization studies using one genetic instrument.
\newblock \emph{International journal of epidemiology}, 42\penalty0 (4):\penalty0 1157--1163, 2013.

\bibitem[Lyall et~al.(2017)Lyall, Celis-Morales, Ward, Iliodromiti, Anderson, Gill, Smith, Ntuk, Mackay, Holmes, et~al.]{lyall2017association}
Donald~M Lyall, Carlos Celis-Morales, Joey Ward, Stamatina Iliodromiti, Jana~J Anderson, Jason~MR Gill, Daniel~J Smith, Uduakobong~Efanga Ntuk, Daniel~F Mackay, Michael~V Holmes, et~al.
\newblock Association of body mass index with cardiometabolic disease in the uk biobank: a mendelian randomization study.
\newblock \emph{JAMA cardiology}, 2\penalty0 (8):\penalty0 882--889, 2017.

\bibitem[Locke et~al.(2015)Locke, Kahali, Berndt, Justice, Pers, Day, Powell, Vedantam, Buchkovich, Yang, et~al.]{locke2015genetic}
Adam~E Locke, Bratati Kahali, Sonja~I Berndt, Anne~E Justice, Tune~H Pers, Felix~R Day, Corey Powell, Sailaja Vedantam, Martin~L Buchkovich, Jian Yang, et~al.
\newblock Genetic studies of body mass index yield new insights for obesity biology.
\newblock \emph{Nature}, 518\penalty0 (7538):\penalty0 197--206, 2015.

\bibitem[Tchetgen~Tchetgen et~al.(2021)Tchetgen~Tchetgen, Sun, and Walter]{tchetgen2021genius}
Eric Tchetgen~Tchetgen, BaoLuo Sun, and Stefan Walter.
\newblock The genius approach to robust mendelian randomization inference.
\newblock \emph{Statistical Science}, 36\penalty0 (3):\penalty0 443--464, 2021.

\bibitem[Bowden and Holmes(2019)]{bowden2019meta}
Jack Bowden and Michael~V Holmes.
\newblock Meta-analysis and mendelian randomization: A review.
\newblock \emph{Research synthesis methods}, 10\penalty0 (4):\penalty0 486--496, 2019.

\bibitem[Lin et~al.(2022{\natexlab{b}})Lin, Seal, and Basu]{lin2022estimating}
Zhaotong Lin, Souvik Seal, and Saonli Basu.
\newblock Estimating snp heritability in presence of population substructure in biobank-scale datasets.
\newblock \emph{Genetics}, 220\penalty0 (4):\penalty0 iyac015, 2022{\natexlab{b}}.

\end{thebibliography}

\section*{Supplemental Information}

\subsection{Proof for $E[A_{ij}e_{ij}^{yx}))]=0$}
\label{GMMproff}
\begin{equation}\label{GMMEq}
    \begin{split}
        E[A_{ij}e_{ij}^{yx}))]=&E[\bm{G_{i}^TG_{j} }(\bm{G_{ic}^T\alpha_{c} }+\bm{G_{id}^T\alpha_{d}}+e_{y_i})(\bm{G_{jb}^T\beta_{b}}+\bm{G_{jc}^T\beta_{c}} +e_j^x)]\\
        \xRightarrow[e_{y_i}\bot e_{x_j}]{e\bot G }&
        E[\bm{G_{i}^TG_{j} }(\bm{G_{ic}^T\alpha_{c} }+\bm{G_{id}^T\alpha_{d}})(\bm{G_{jb}^T\beta_{b}}+\bm{G_{jc}^T\beta_{c}})]\\
        \xRightarrow[e_{y_i}\bot e_{x_j}, e\bot G]{\bm{\mathbb{G}_a\bot \mathbb{G}_b\bot \mathbb{G}_c\bot \mathbb{G}_d} }&
        E[\bm{G_{i}^TG_{j} }(\bm{G_{ic}^T\alpha_{c} }+\bm{G_{id}^T\alpha_{d}})(\bm{G_{jb}^T\beta_{b}}+\bm{G_{jc}^T\beta_{c}})]\\
        \xRightarrow[\alpha_{c}\bot \beta_{b}]{\alpha_{d}\bot \beta_{b},\beta_{c} }&
        E[\bm{G_{i}^TG_{j} }\bm{G_{ic}^T\alpha_{c} }\bm{G_{jc}^T\beta_{c}}]\\
        \xRightarrow[]{E(\alpha_{c}\beta_{c})}&
        0.
    \end{split}
\end{equation}

\subsection{Bias and asymptotic variance of the TS-RE}
\label{BiasProof}
 In the following proof, we used $r$ to denote terms with expectation $0$ ($E(r)=0$). Here the genotype data are assumed standardized that $E(G_{ik}=0)$,$ Var(G_{ik})=1, E(A_{ij}=0)$. Let $\delta=E\left[A_{ij}X_iX_j\right], \eta=E\left[A_{ij}X_iY_j\right]$, then $\hat{\theta}_{TS-RE}$ can be regarded as a ratio of $\eta/\delta$.
 \begin{equation}\label{delta}
  \begin{split}
      \hat{\delta}  =&\frac{1}{N} \sum_{i < j} E(A_{ij}X_iX_j)\\
      \xRightarrow[e_{x_j}\bot Y_i]{e_{y_i}\bot X_j}&E[A_{ij} (\bm{G_{ib}^T\beta_{b}}+\bm{G_{ic}^T\beta_{c}})(\bm{G_{jb}^T\beta_{b}}+\bm{G_{jc}^T\beta_{c}})+r]\\
       \xRightarrow[p\neq q]{\beta{p}\bot \beta_{q} }&\frac{1}{M} E[\sum_{k=1}^M G_{ik} G_{jk}(\bm{G_{ib}^TG_{jb}}\beta^2_{bk}+\bm{G_{ic}^TG_{jc}}\beta^2_{ck})]\\
       \xRightarrow[p\neq q]{G_{.p}\bot G_{. q} }&\frac{1}{M} E[\sum_{k=1}^{M_b} G_{ik}^2 G_{jk}^2\beta^2_{bk}+\sum_{k=1}^{M_c} G_{ik}^2 G_{jk}^2\beta^2_{ck}]\\
        =&  \frac{M_b}{M}E(\beta_{bk}
        ^2)+\frac{M_c}{M}E(\beta_{ck}
        ^2)
  \end{split}
 \end{equation}

Similarly, using Equation \ref{GMMEq} and Equation \ref{delta}
\begin{equation}
  \begin{split}
    \hat{\eta} =& E(\widehat{Cov}\left[A_{ij},Y_iX_j\right])\\
      =& \frac{1}{M} E[\theta\sum_{k=1}^{M_b} G_{ik}^2 G_{jk}^2\beta^2_{bk}+\sum_{k=1}^{M_c} G_{ik}^2 G_{jk}^2(\theta \beta^2_{ck}+\beta_{ck}\alpha_{ck}))]\\
      =&\frac{M_b}{M}\theta E(\beta_{bk}
        ^2)+\frac{M_c}{M}\theta[E(\beta_{ck}
        ^2)+E(\beta_{ck}\alpha_{ck})]
  \end{split}
 \end{equation}
 The ratio of $\hat{\eta}$ and $\hat{\delta}$ is the TS-RE estimator
 \begin{equation}
     \begin{split}
\hat{\theta}_{TS-RE}=\frac{\hat{\eta}}{\hat{\delta}}=\theta+\frac{M_cE(\beta_{ck}\alpha_{ck})}{M_bE(\beta_{bk}^2)+M_cE(\beta_{ck}
        ^2)}. 
     \end{split}
 \end{equation}
The bias will tend to zero if $E(\beta_c\alpha_c)=0$, e.g., the Inside assumption holds, $\rho_{\mathbb{G}_c}=0$, and either  $\mu_{\mathbb{G}_c^x}=0$ or $\mu_{\mathbb{G}_c^y}=0$, t. Given the $\theta=\frac{\eta}{\delta}$, we can use the multivariate Delta method to derive the asymptotic property for $\hat{\theta}_{GMM}$,
 $$\sqrt{N}[\hat{\theta}_{TS-RE}-\theta] \xrightarrow[]{\mathcal{D}} N(0, \tau^2),$$
 where 
 \begin{equation}
     \begin{split}
         \tau^2 =&\frac{\eta^2}{\delta^2}(\frac{Var(A_{ij}Y_iX_j)}{\eta^2}+\frac{Var(A_{ij}X_iX_j)}{\delta^2}-2\frac{Cov(A_{ij}Y_iX_j,A_{ij}X_iX_j)}{\eta\delta})\\
         =&\frac{M[M_bE(\beta_{bk}
        ^2)+M_cE(\beta_{ck}
        ^2)+\sigma_{e_x}^2][M_cE(\alpha_{ck}
        ^2)+M_dE(\alpha_{dk}
        ^2)+\sigma_{e_y}^2]}{[M_bE(\beta_{bk}
        ^2)+M_cE(\beta_{ck}
        ^2)]^2}
     \end{split}
 \end{equation}
 Thus
$Var(\hat{\theta}_{TS-RE})=\frac{2M}{n(n-1)}\frac{[M_bE(\beta_{bk}
        ^2)+M_cE(\beta_{ck}
        ^2)+\sigma_{e_x}^2][M_cE(\alpha_{ck}
        ^2)+M_dE(\alpha_{dk}
        ^2)+\sigma_{e_y}^2]}{[M_bE(\beta_{bk}
        ^2)+M_cE(\beta_{ck}
        ^2)]^2}$.
To show the relationship between our method with IVW and Egger, we consider the situation that all included IVs have a direct effect on $X$, which are from $\mathbb{G}_b$ and $\mathbb{G}_c$, then
 \begin{equation}\label{BiasIVW}
    \begin{split}
       E(\hat{\theta}_{IVW})
       &=  E(\bm{(X^TP_GX)^{-1}X^TP_GY})\\
        &\xRightarrow[G_{ik}^2=1]{\beta{p}\bot \beta_{q}, p\neq q} \theta+\frac{M_cE(\beta_{ck}\alpha_{ck})}{M_bE(\beta_{bk}^2)+M_cE(\beta_{ck}
        ^2)}
    \end{split}
\end{equation} 
 
 \begin{equation}\label{BiasEgger}
    \begin{split}
       E(\hat{\theta}_{Egger})
      &= E(\frac{\bm{I^TG^TGIX^TP_GY-X^TGIY^TGI}}{\bm{I^TG^TGIX^TP_GX-X^TGIX^TGI}})\\
     & \xRightarrow[G_{ik}^2=1]{\beta{p}\bot \beta_{q}, p\neq q} \theta + \frac{M_c}{M} \frac{E(\beta_{ck}\alpha_{ck})-E(\beta_{ck})E(\alpha_{ck})}{Var(\beta)}
    \end{split}
\end{equation} 
Thus, if there are only two groups of IVs $\mathbb{G_b},\mathbb{G_c}$, the bias of TS-RE is equivalent to the IVW.

\newpage
\subsection{Simulation results}
\subsubsection{Tables for simulation investigating IVs from $\mathbb{G}_b$: impact of IVs numbers and Hertability}
\begin{table}[!ht]
\renewcommand\thetable{S1} 
\tiny
    \centering
    \begin{tabular}{cccccccccccc}
    \hline
       $M_b$ &  $\sigma_g$& Her& IVs & SM & WM & IVW & Egger& Lasso & dIVW & RAPS &TS-RE\\
       \hline
  \multirow{4}{*}{$100$} 
      &0.03&0.05& All &  0.52 (0.11) & 0.53 (0.09) & 0.53 (0.07) & 0.53 (0.12) &0.54 (0.06)& 2.04 (1.33)& 0.77 (0.51)&0.30 (0.86)\\
    & &&Top20 & 0.53 (0.12) & 0.54 (0.11)  & 0.54 (0.09) & 0.59 (0.40)&0.54 (0.08)&0.70 (0.11) & 0.59 (0.16) &0.50 (0.14)\\
  \cline{2-12}
      &0.05&0.11& All &  0.44 (0.08) & 0.44 (0.07) & 0.44 (0.05) & 0.45 (0.10)& 0.44 (0.06)&0.87 (0.16)& 0.59 (0.17) & 0.27 (0.18)\\
    & &&Top20 & 0.44 (0.08) & 0.45 (0.09)  & 0.53 (0.34) & 0.54 (0.07) &0.45 (0.07)&0.53 (0.08)& 0.48 (0.11)& 0.42 (0.09)\\    
     \hline
  \multirow{4}{*}{$1000$}  
   &0.03&0.31& All &  0.49 (0.04)  &  0.49 (0.03)   & 0.49 (0.03)  &0.48 (0.04) &0.48 (0.03)&2.08 (0.32) & 0.72(0.17)&0.28 (0.12)\\
          &&& Top20 &   0.47 (0.07)  &  0.48 (0.07)   & 0.48 (0.05)  &0.55 (0.54) &0.48 (0.06)& 0.54 (0.06)&0.51 (0.08)& 0.47 (0.07)\\
       \cline{2-12}
            &0.05&0.56& All &  0.40 (0.03) & 0.40 (0.02) & 0.40 (0.03) & 0.40 (0.03) &0.40 (0.02)&1.13 (0.10)& 0.55 (0.07)&0.29 (0.05)\\
    & &&Top20 & 0.40 (0.05) & 0.40 (0.05) & 0.40 (0.04) & 0.35 (0.36)&0.40 (0.04)&0.44 (0.05) &0.42 (0.06)& 0.40 (0.05)\\
     \hline
  \multirow{6}{*}{$5000$}  
            &0.03&0.69& All  & 0.40 (0.02) &  0.40 (0.02) &  0.40 (0.02) &  0.40 (0.02)&0.40 (0.02)& 3.34 (0.51)& 0.58 (0.11)&0.30 (0.07)\\
    & &&Top20&  0.40 (0.04) &  0.40 (0.04) &  0.40 (0.04) &  0.36(0.43) & 0.40 (0.04)&0.44 (0.04)& 0.41 (0.05)&0.40 (0.05)\\
  \cline{2-12}
             &0.05&0.86& All  & 0.34 (0.01) & 0.34 (0.01)  & 0.34 (0.01)  & 0.34 (0.01)  &0.34 (0.01)&2.39 (0.24)& 0.44 (0.06)&0.30 (0.04)\\
    & &&Top20& 0.34 (0.03) &0.34 (0.03) & 0.34 (0.02) & 0.32 (0.30) & 0.38 (0.02)&0.38 (0.03)&0.34 (0.04)&0.34 (0.03)\\
  \cline{2-12}
          &0.05&0.42& All &  0.50 (0.03)  & 0.50 (0.03)  &0.50 (0.02)  &0.50 (0.03)  &0.50 (0.02)&6.94 (1.82)& 0.55 (0.12)& 0.30 (0.16)\\
         & &&Top20 &  0.51 (0.06)  & 0.51 (0.06)  &0.51 (0.02)  &0.48 (0.61) &0.51 (0.05)&0.56 (0.05) &0.54 (0.07) &0.51 (0.07)\\
     \hline
    \end{tabular}
    \caption{Mean and SE of different methods:  impact of number of IVs and genetic variance. The sample size is set to be 1000. For each method, we used all SNPs as IVs or selected top 20 significant IVs.}
    \label{Table41}
    \end{table}

\begin{table}[!ht]
\renewcommand\thetable{S2} 
\tiny
    \centering
    \begin{tabular}{ccccccccccc}
    \hline
       $M_b$ & $p_w$ & IVs & SM & WM & IVW & Egger&Lasso& dIVW & RAPS& TS-RE\\
            \hline
 \multirow{4}{*}{100}   &  \multirow{2}{*}{0.8}& All & 0.36 (0.06)   &0.34 (0.04)  & 0.35 (0.04) & 0.33 (0.05) &0.35 (0.04)   &0.45 (0.05) & 0.40 (0.06)& 0.29 (0.06)\\
                          &   & Top20 &  0.33 (0.05)  & 0.33 (0.05) & 0.33 (0.04) &0.33 (0.17) &  0.33 (0.04) & 0.36 (0.04)& 0.35 (0.05)&0.32 (0.05)\\
       \cline{2-11}
        &  \multirow{2}{*}{1}   & All & 0.44 (0.08) & 0.44 (0.07) & 0.44 (0.05) & 0.45 (0.10) & 0.44 (0.06)& 0.87 (0.16)  & 0.59 (0.17) &0.27 (0.17)\\
 &   & Top20 & 0.44 (0.08) & 0.45 (0.09) & 0.45 (0.06) & 0.53 (0.34)  &0.45 (0.07)& 0.53 (0.08) &0.48 (0.11) & 0.42 (0.09)\\
      
 \hline
  \multirow{4}{*}{$1000$}   &  \multirow{2}{*}{0.8}& All &  0.33 (0.02)  &  0.33 (0.02) & 0.33 (0.01)  & 0.33 (0.02)  & 0.33 (0.1)  & 0.73 (0.04)& 0.40 (0.03)&0.30 (0.02)\\
       &   & Top20 &   0.33 (0.03)  &   0.33 (0.03)&  0.33 (0.02) &  0.33 (0.20)& 0.33 (0.02)  & 0.36 (0.03) &0.32 (0.04) & 0.33 (0.03)\\
       \cline{2-11}
        &  \multirow{2}{*}{1}   & All & 0.40 (0.03) & 0.40 (0.02) & 0.40 (0.02) & 0.40 (0.03) & 0.40 (0.02)& 1.13 (0.10) & 0.55 (0.07) & 0.29 (0.05)\\
 &   & Top20 & 0.40 (0.05) & 0.40 (0.05) & 0.40 (0.04) & 0.35 (0.36) & 0.40 (0.04)  & 0.44 (0.05)&0.42 (0.06)& 0.40 (0.05)\\
     \hline
  \multirow{4}{*}{$5000$}   &  \multirow{2}{*}{0.8}& All &  0.31 (0.01)  & 0.31 (0.01) & 0.31 (0.01) &0.31 (0.01)  & 0.31 (0.01)  &1.93 (0.19)&0.34 (0.03) & 0.30 (0.02)\\
       &   & Top20 &  0.31 (0.01)  & 0.31 (0.01) & 0.31 (0.01) & 0.32 (0.14)& 0.31 (0.01)  & 0.34 (0.01) &0.31 (0.03) &0.31 (0.01)\\
       \cline{2-11}
        &  \multirow{2}{*}{1}   & All & 0.34 (0.01) & 0.34 (0.01) & 0.34 (0.01) & 0.34 (0.01)&  0.34 (0.01) & 2.39 (0.24)&  0.44 (0.06)&0.30 (0.04)\\
 &   & Top20 & 0.34 (0.03) & 0.34 (0.03) & 0.34 (0.02) & 0.33 (0.30) &0.35 (0.02)& 0.38 (0.03)  & 0.34 (0.04) & 0.34 (0.03)\\
     \hline
    \end{tabular}
    \caption{Mean and SE of different methods: $20\%$ of the IVs having strong effects on $X$. The large effect for $20\%$ of the IVs is $0.2$. The variance parameter $\sigma_{\mathbb{G}_b}$ was $0.05$ and the corresponding heritability values are $0.11, 0.56, 0.86$. The true causal effect is $\theta=0.3$}
    \label{Table42}
\end{table}

\subsubsection{IVs from $\mathbb{G}_b$ under large sample sizes}
Given the necessity for large sample sizes in other MR methods, we conducted an investigation involving different sample sizes ranging from 1000 to 10000. We held certain parameters constant, specifically setting $M_b=1000$, $\sigma_{g_b}=0.03$, and a causal effect of $\theta=0.3$. The residual variance parameter $\sigma_{e_x}^2$ was maintained at a fixed value of 2, and the Her remained at 0.31. 

Table S3 includes more detailed results. In cases where all IVs lacked strength ($\mu_{g_b}=0$), it was evident that enlarging the sample size led to some reduction in bias for other MR methods. Nevertheless, these methods still yielded biased estimates. In stark contrast, our TS-RE consistently produced unbiased estimates.

Under configurations where $20\%$ of the IVs possessed strong effects ($\mu_{g_b}=0.2$), our TS-RE estimates remained unbiased across all sample sizes. In contrast, the other MR methods required sample sizes of $n\geq 5000$ to achieve unbiased estimates. Notably, when the top 20 significant IVs were used, Egger exhibited the largest standard error SE compared to the other methods. For the weak-IV MR method dIVW, increasing the sample size significantly reduced bias, but it still displayed the largest bias when all IVs were included. This may be attributed to the fact that the consistency of the dIVW estimator relies on a very large "effective sample size" as defined in \citep{ye2021debiased}, a condition not guaranteed in our simulation setting.
\begin{table}[!ht]
\renewcommand\thetable{S3} 
    \centering
    \tiny
     \begin{tabular}{ccccccccccc}
    \hline
       n &$p_w$& IVs & SM & WM & IVW & Egger& Lasso& dIVW& RAPS&TS-RE\\
       \hline
    \multirow{4}{*}{1000} & \multirow{2}{*}{0.8}& All &  0.33 (0.02)  &  0.33 (0.02)   & 0.33 (0.01)  &0.33 (0.02) & 0.34 (0.01)  & 0.75 (0.05)&0.41 (04) & 0.30 (0.02)\\
    &   & Top20 & 0.33 (0.03) & 0.33 (0.03) & 0.33 (0.02) & 0.35 (0.20)&  0.33 (0.03) & 0.36 (0.02) & 0.34 (0.04)&0.33 (0.03)\\
     \cline{2-11}
      & \multirow{2}{*}{1}& All &  0.49 (0.04)  &  0.49 (0.03)   & 0.49 (0.03)  &0.48 (0.04) & 0.48 (0.03)  &  2.01 (0.32)&0.72 (0.17)& 0.28 (0.12)\\
      &   & Top20 & 0.47 (0.07) & 0.48 (0.07) & 0.48 (0.05) & 0.55 (0.54) &0.47 (0.06)& 0.54 (0.06) &0.51 (0.08) &  0.47 (0.07)\\
       \hline
         \multirow{4}{*}{3000} & \multirow{2}{*}{0.8}& All &  0.33 (0.01)  &  0.31 (0.01)   & 0.32 (0.01)  &0.31 (0.01) & 0.32 (0.01)  &0.45 (0.01)&0.36 (0.02)  & 0.30 (0.01)\\
         &   & Top20 & 0.32 (0.02) & 0.32 (0.02) & 0.32 (0.02) & 0.32 (0.20) &   0.32 (0.02) & 0.33 (0.02) &0.32 (0.03) & 0.32 (0.02)\\
          \cline{2-11}
      & \multirow{2}{*}{1}& All &  0.42 (0.03)  &  0.42 (0.02)   & 0.42 (0.02)  &0.42 (0.03)  & 0.42 (0.02)  &0.88 (0.07)& 0.57 (0.05)& 0.30 (0.05)\\
      &   & Top20 & 0.42 (0.05) & 0.42 (0.05) & 0.42 (0.04) & 0.43 (0.41) &0.42 (0.04)&  0.45 (0.04) &  0.44 (0.06)& 0.42 (0.06)\\
       \hline
         \multirow{4}{*}{5000} & \multirow{2}{*}{0.8}& All &  0.32 (0.01)  &  0.31 (0.01)   & 0.31 (0.01)  &0.30 (0.01)  &  0.31 (0.01) &0.39 (0.01)&0.34 (0.01) & 0.30 (0.01)\\
         &   & Top20 & 0.31 (0.01) & 0.31 (0.01) & 0.31 (0.01) & 0.33 (0.20) &0.31 (0.01)& 0.32 (0.01)  & 0.32 (0.02)&  0.31 (0.02)\\
          \cline{2-11}
      & \multirow{2}{*}{1}& All &  0.39 (0.02)  &  0.39 (0.02)   & 0.39 (0.01)  &0.39 (0.02)  & 0.39 (0.02)&  0.64 (0.03)& 0.50 (0.05)& 0.30 (0.02)\\
      &   & Top20 & 0.39 (0.04) & 0.39 (0.05) & 0.39 (0.04) & 0.34 (0.37) &0.38 (0.03)&  0.41 (0.04) &  0.41 (0.05)&0.38 (0.05)\\
       \hline
         \multirow{4}{*}{10000} &\multirow{2}{*}{0.8}& All &  0.31 (0.007)  &  0.30 (0.006)   & 0.31 (0.004)  &0.30 (0.005) & 0.30 (0.005)  &0.30 (0.005) &0.32 (0.007) & 0.30 (0.005)\\
         &   & Top20 & 0.31 (0.01) & 0.31 (0.01) & 0.31 (0.01) & 0.36 (0.15) &0.30 (0.01)   & 0.31 (0.01)&0.31 (0.02)& 0.31 (0.01)\\
          \cline{2-11}
      & \multirow{2}{*}{1}& All &  0.36 (0.02)  &  0.36 (0.02)   & 0.36 (0.01)  &0.36 (0.02) &  0.36 (0.01) &0.47 (0.02) &0.42 (0.02)  & 0.30 (0.02)\\
      &   & Top20 & 0.36 (0.04) & 0.36 (0.05) & 0.36(0.04) & 0.35 (0.29) &0.35 (0.03)& 0.37 (0.03)&0.37 (0.04)  &  0.35 (0.04)\\
       \hline
    \end{tabular}
    \caption{Mean and SE of different methods under different sample sizes, $M_b=1000$}
    \label{Table43}
\end{table}
\hspace*{-2cm}

\newpage
\subsubsection{Pleiotropic IVs from $\mathbb{G}_c$: different proportion of valid IVs}
In our investigation, we explored the influence of the proportion of valid IVs on the estimates while maintaining a fixed sample size of $n=1000$ and a total of $M=1000$ IVs. We varied the number of valid IVs, denoted as $M_b$, ranging from 0 to 1000. The causal effect was set to be $\theta=0.1, 0.3$. The genetic variances were set to $\sigma_{g_b^x}=\sigma_{g_c^x}=\sigma_{g_c^y}=0.03$, and residual variances were $\sigma_{e_x}^2=\sigma_{e_y}^2=2$. For both balanced pleiotropy ($E(\alpha_c)=0$) and directional pleiotropy ($E(\alpha_c)=0.1$), we ensured that the InSIDE assumption remained valid. All IVs were considered weak ($E(\beta_b)=E(\beta_c)=0$), and the heritability was set to $Her=0.31$. In Figure~\ref{FigS1}, solid lines represent bias, while dashed lines represent SE for each method.

\begin{figure}[!ht]
\centering
\renewcommand\thefigure{S1} 
\includegraphics[width=\textwidth]{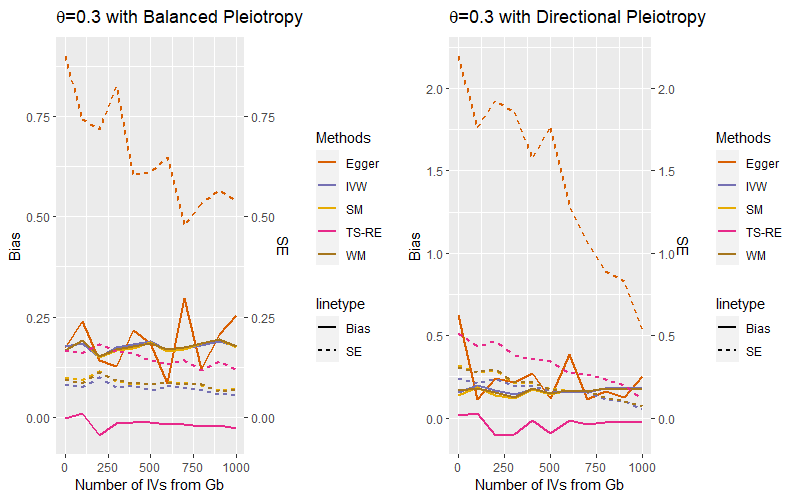}
\caption{ Bias and standard error (SE) of the estimates the causal effect $\theta=0.3$, the solid lines are biases and dashed lines are SEs. For the balanced pleiotropy $E(\alpha_c)=0$ and for the directional pleiotropy $E(\alpha_c)=0.1$, the InSIDE assumption is valid. Here total number of IVs is $M=1000$, the sample size is $n=1000$, $E(\beta_b)=E(\beta_c)=0$ and $\sigma_{\mathbb{G}_b}=\sigma_{\mathbb{G}_c}=0.03$, $Her=0.31$. TS-RE used all IVs while other MR methods used the selected top 20 most significant IVs.}
\label{FigS2}
\end{figure}

When all IVs were used, TS-RE consistently exhibited much smaller biases compared to other methods, regardless of whether pleiotropy was balanced or directional. Selecting the top 20 significant IVs substantially increased the bias and SE for the TS-RE method.

Effect of Proportion of Invalid IVs on SE with Directional Pleiotropy: Under directional pleiotropy ($E(\alpha_c)=0.1$), a higher proportion of invalid IVs resulted in higher SE for the TS-RE estimator. This was primarily due to the increased contribution of IVs from $\mathbb{G}_c$, which significantly increased the value of the term $M_cE(\alpha_k^2)=M_cE(\alpha_c)^2+M_cVar(\alpha_c)$ in the numerator of $\tau^2$. Notably, TS-RE exhibited a larger bias when a moderate proportion of IVs were valid (around $30-50\%$).

Effect of Proportion of Invalid IVs on SE with Balanced Pleiotropy: In contrast, under balanced pleiotropy ($E(\alpha_c)=0$), the impact of the proportion of valid IVs on SE was considerably reduced compared to the scenario with directional pleiotropy. This was because, for balanced pleiotropy, the value $M_cE(\alpha_k^2)=0.03^2M_c$ was much smaller than the value for directional pleiotropy, where $M_cE(\alpha_k^2)=(0.1^2+0.03^2)M_c$.

In summary, TS-RE consistently exhibited lower bias than other methods when all IVs were used, regardless of the type of pleiotropy. However, the choice to select only the top 20 significant IVs for TS-RE significantly increased bias and SE. The impact of the proportion of invalid IVs on SE was more pronounced under directional pleiotropy compared to balanced pleiotropy, where the effect on SE was mitigated.

\begin{table}[!ht]
\renewcommand\thetable{S4} 
\tiny
    \centering
    \begin{tabular}{cccccccccccccc}
    \hline
$M_b$ &  Method &\multicolumn {6}{c}{Balanced Pleiropy} &\multicolumn {6}{c}{Directional Pleiropy}\\
\cline{3-14}
&&  \multicolumn {3}{c}{All IVs} &\multicolumn {3}{c}{Top 20 IVs}& \multicolumn {3}{c}{All IVs} &\multicolumn {3}{c}{Top 20 IVs}\\
  \hline
  &&Bias&SE&MSE&Bias&SE&MSE&Bias&SE&MSE&Bias&SE&MSE\\
  \hline
\multirow {5}{*}{0}&SM&0.18&0.06&0.04&0.17&0.1&0.04&0.23&0.19&0.09&0.14&0.32&0.12\\
&WM&0.18&0.06&0.04&0.17&0.09&0.04&0.22&0.17&0.08&0.17&0.31&0.12\\
&IVW&0.18&0.05&0.04&0.18&0.08&0.04&0.2&0.14&0.06&0.15&0.24&0.08\\
&Egger&0.18&0.07&0.04&0.17&0.9&0.84&0.19&0.16&0.06&0.62&2.19&5.21\\
&TS-RE&0&0.16&0.03&0.17&0.11&0.04&0.01&0.51&0.26&0.14&0.29&0.11\\
  \hline
\multirow {5}{*}{100}&SM&0.19&0.06&0.04&0.19&0.09&0.04&0.21&0.19&0.08&0.19&0.27&0.11\\
&WM&0.19&0.05&0.04&0.19&0.08&0.04&0.2&0.16&0.07&0.18&0.28&0.11\\
&IVW&0.19&0.05&0.04&0.18&0.07&0.04&0.21&0.12&0.06&0.2&0.21&0.08\\
&Egger&0.19&0.06&0.04&0.24&0.74&0.61&0.2&0.16&0.06&0.12&1.76&3.12\\
&TS-RE&0.01&0.16&0.03&0.18&0.1&0.04&0.03&0.43&0.19&0.2&0.27&0.11\\
  \hline
\multirow {5}{*}{200}&SM&0.15&0.08&0.03&0.15&0.12&0.04&0.18&0.16&0.06&0.14&0.29&0.1\\
&WM&0.15&0.08&0.03&0.15&0.11&0.04&0.17&0.14&0.05&0.16&0.3&0.11\\
&IVW&0.16&0.07&0.03&0.15&0.1&0.03&0.16&0.12&0.04&0.17&0.24&0.09\\
&Egger&0.15&0.08&0.03&0.14&0.72&0.53&0.14&0.15&0.04&0.24&1.92&3.76\\
&TS-RE&-0.05&0.18&0.03&0.13&0.12&0.03&-0.11&0.47&0.23&0.17&0.29&0.11\\
  \hline
\multirow {5}{*}{300}&SM&0.18&0.05&0.03&0.17&0.09&0.04&0.17&0.14&0.05&0.12&0.21&0.06\\
&WM&0.18&0.05&0.03&0.17&0.09&0.04&0.15&0.12&0.04&0.13&0.21&0.06\\
&IVW&0.18&0.04&0.03&0.17&0.08&0.04&0.17&0.12&0.04&0.14&0.19&0.06\\
&Egger&0.18&0.06&0.04&0.13&0.82&0.7&0.15&0.14&0.04&0.22&1.86&3.5\\
&TS-RE&-0.02&0.16&0.03&0.16&0.1&0.04&-0.1&0.38&0.15&0.12&0.23&0.07\\
  \hline
\multirow {5}{*}{400}&SM&0.18&0.05&0.04&0.17&0.08&0.04&0.18&0.14&0.05&0.18&0.21&0.08\\
&WM&0.18&0.05&0.03&0.18&0.08&0.04&0.19&0.12&0.05&0.18&0.22&0.08\\
&IVW&0.18&0.04&0.03&0.18&0.08&0.04&0.19&0.11&0.05&0.18&0.19&0.07\\
&Egger&0.18&0.06&0.04&0.22&0.6&0.41&0.19&0.13&0.05&0.27&1.57&2.55\\
&TS-RE&-0.01&0.16&0.03&0.18&0.1&0.04&-0.01&0.36&0.13&0.18&0.23&0.08\\
  \hline
\multirow {5}{*}{500}&SM&0.18&0.05&0.04&0.19&0.08&0.04&0.16&0.11&0.04&0.15&0.18&0.05\\
&WM&0.19&0.05&0.04&0.18&0.08&0.04&0.17&0.11&0.04&0.15&0.17&0.05\\
&IVW&0.19&0.04&0.04&0.19&0.07&0.04&0.17&0.09&0.04&0.15&0.17&0.05\\
&Egger&0.19&0.05&0.04&0.18&0.61&0.41&0.16&0.12&0.04&0.12&1.77&3.14\\
&TS-RE&-0.01&0.14&0.02&0.18&0.08&0.04&-0.1&0.34&0.13&0.14&0.2&0.06\\
  \hline
\multirow {5}{*}{600}&SM&0.19&0.05&0.04&0.17&0.09&0.04&0.19&0.1&0.05&0.17&0.16&0.05\\
&WM&0.18&0.05&0.03&0.17&0.09&0.04&0.19&0.1&0.04&0.17&0.16&0.05\\
&IVW&0.18&0.04&0.04&0.17&0.08&0.03&0.19&0.09&0.04&0.16&0.16&0.05\\
&Egger&0.18&0.06&0.03&0.08&0.65&0.43&0.17&0.11&0.04&0.39&1.29&1.81\\
&TS-RE&-0.02&0.13&0.02&0.15&0.1&0.03&-0.01&0.27&0.07&0.15&0.18&0.06\\
  \hline
\multirow {5}{*}{700}&SM&0.19&0.05&0.04&0.17&0.08&0.04&0.18&0.09&0.04&0.16&0.16&0.05\\
&WM&0.18&0.05&0.04&0.17&0.08&0.04&0.17&0.09&0.04&0.16&0.16&0.05\\
&IVW&0.19&0.04&0.04&0.17&0.07&0.04&0.18&0.08&0.04&0.16&0.15&0.05\\
&Egger&0.18&0.05&0.04&0.3&0.48&0.32&0.16&0.11&0.04&0.11&1.06&1.14\\
&TS-RE&-0.02&0.14&0.02&0.16&0.09&0.03&-0.04&0.26&0.07&0.14&0.19&0.05\\
  \hline
\multirow {5}{*}{800}&SM&0.18&0.04&0.04&0.18&0.08&0.04&0.19&0.08&0.04&0.18&0.12&0.05\\
&WM&0.18&0.04&0.04&0.18&0.08&0.04&0.18&0.07&0.04&0.18&0.12&0.05\\
&IVW&0.18&0.04&0.03&0.18&0.07&0.04&0.18&0.07&0.04&0.18&0.11&0.05\\
&Egger&0.18&0.05&0.04&0.12&0.53&0.3&0.18&0.09&0.04&0.16&0.89&0.81\\
&TS-RE&-0.02&0.12&0.01&0.16&0.09&0.03&-0.03&0.23&0.05&0.17&0.14&0.05\\
  \hline
\multirow {5}{*}{900}&SM&0.19&0.04&0.04&0.19&0.07&0.04&0.19&0.07&0.04&0.18&0.1&0.04\\
&WM&0.19&0.04&0.04&0.19&0.06&0.04&0.18&0.05&0.04&0.18&0.1&0.04\\
&IVW&0.19&0.03&0.04&0.19&0.06&0.04&0.19&0.06&0.04&0.19&0.1&0.04\\
&Egger&0.18&0.04&0.04&0.21&0.56&0.36&0.18&0.06&0.04&0.12&0.83&0.71\\
&TS-RE&-0.02&0.14&0.02&0.18&0.08&0.04&-0.02&0.2&0.04&0.18&0.13&0.05\\
  \hline
\multirow {5}{*}{1000}&SM&0.19&0.04&0.04&0.17&0.07&0.03&0.19&0.04&0.04&0.17&0.07&0.03\\
&WM&0.18&0.03&0.03&0.18&0.07&0.04&0.18&0.03&0.03&0.18&0.07&0.04\\
&IVW&0.18&0.03&0.03&0.18&0.05&0.03&0.18&0.03&0.03&0.18&0.05&0.03\\
&Egger&0.18&0.04&0.03&0.25&0.54&0.35&0.18&0.04&0.03&0.25&0.54&0.35\\
&TS-RE&-0.03&0.12&0.01&0.17&0.07&0.03&-0.03&0.12&0.01&0.17&0.07&0.03\\
\hline
    \end{tabular}
    \caption{Simulation results for a mixture of IVs from $\mathbb{G}_b$ and $\mathbb{G}_c$. The total number of IVs is $1000=M_b+M_c$ and the number $M_b$ is varied from $0$ to $1000$. For the balanced pleiotropy $E(\alpha_c) = 0$, and for the directional pleiotropy $E(\alpha_c) = 0.1$, the InSIDE assumption is valid that $\rho_{\mathbb{G}_c}=0$. All IVs have weak effect $N(\mu=0, \sigma^3=0.03^3)$ and the true causal effect is $\theta=0.3$}
    \label{S4}
\end{table}
\end{document}